\documentstyle[10pt,epsfig]{article}
\begin{document}

\renewcommand{\thefootnote}{\fnsymbol{footnote}}
\begin{center}
{\LARGE Form Factors of Meson Decays \\in the Relativistic Constituent Quark
Model}
\vspace{0.5cm}

{\Large Dmitri Melikhov}
\\
\vspace{0.2cm}

{\normalsize {\it Nuclear Physics Institute, Moscow State University, Moscow,
119899, Russia \footnote{E-mail: melikhov@sgi.npi.msu.su}}}
\end{center}

A formalism for the relativistic description of hadron decays
within the constituent quark model is presented.
First, hadron amplitudes of the light--cone constituent quark model, in
particular the weak transition
form factors at spacelike momentum transfers, $q^2\le 0$, are represented in
the form of the dispersion integrals over the
hadron mass.
Second, the form factors at $q^2>0$ are obtained by performing the analytic
continuation from the region $q^2<0$. As a result, the transition form factors
both in the
scattering and the decay regions are
expressed through light--cone wave functions of the initial and final hadrons.
The technique is applied to the description of the semileptonic decays
of pseudoscalar mesons and direct calculation of the transition form factors at
$q^2>0$.

\section{Introduction}
Weak decays of hadrons provide an important source of information on
the parameters of the standard model of electroweak interactions, the structure
of
weak currents,
and internal structure of hadrons. Hadron decay rates involve both the
Cabibbo--Kobayashi--Maskawa matrix elements and hadron form factors,
therefore the extraction of the standard model parameters from the experiments
on hadron decays requires reliable information on hadron structure.

The problem of theory is to describe hadron form factors which involve both
perturbative and nonperturbative contributions. Higher order corrections to
weak
currents are calculable perturbatively and can be predicted to high accuracy.
The calculation of hadronic matrix elements of the weak currents inevitably
encounters the problem of describing the hadron structure and requires
the nonperturbative consideration. This gives the main uncertainty to the
theoretical
predictions for the hadron transition amplitudes.

In the case of the semileptonic $K_{l3}$ decay, the $K\to \pi$ weak transition
form factor deviates from unity only at the second order in comparatively
small $SU(3)$--symmetry breaking and can be calculated to high accuracy
\cite{lr},
that provides the most accurate value of the $V_{us}$.
For extracting the $V_{cd}$, $V_{cs}$, $V_{ub}$, and $V_{cb}$
from the decay rates and lepton spectra in $D_{l3}$ and $B_{l3}$ decays, a
reliable calculation of hadron transition form factors at timelike momentum
transfers
is necessary.

In last decade an increasing amount of publications have been devoted both to
the perturbative calculation of higher order corrections to weak currents
and to the description of weak matrix elements of hadrons. We shall concentrate
on the
latter problem, closely related to the investigation of the nonperturbative
aspect of
hadron structure.

Various theoretical approaches have been applied to the calculation of the
nonperturbative contribution to hadron form factors.
The most popular amond them are the
quark model \cite{wsb}--\cite{card}, QCD sum rules \cite{sr1}--\cite{sr6}, and
lattice QCD calculations \cite{lat3}--\cite{lat5}.
A critical comparison of these approaches can be found, e.g. in \cite{aley}.

Systems containing heavy and light quarks are usually considered within the
Heavy Quark Effective Theory (HQET) \cite{hqet},
an effective theory based on QCD in the limit of infinitely large quark masses.
The $B\to D,D^*$ decays associated with the
heavy-to-heavy $(b\to c)$ quark transition are described in terms of a single
universal Isgur--Wise (IW) function \cite{iw} which can be estimated with any
of the mentioned
nonperturbative approaches. The $O(1/m^N_Q)$ corrections to this picture can be
consistently calculated within the HQET (for a detailed review see
\cite{neubert}).

For the decays caused by the heavy-to-light quark transitions $(D\to K,K^*;
D\to\pi,\rho;$ and $B\to \pi,\rho)$ the situation turns out to be less
definite. The HQET
does not work properly in this case, and theory faces at least two practical
problems, namely:
(i) The existing theoretical considerations fail to describe the experimental
results
for the $B\to K^*\psi$ decay \cite{aley}; (ii) In the absense of experimental
information on $B\to \pi,\rho$ decay modes, the uncertainty of the theoretical
predictions for relevant form factors is too large to make any definite
conclusion on their
values (see Table \ref{table:b2pi}).

This stimulates further investigation of hadron transition form factors.

Our special interest lies in the quark model which reflects at the
phenomenological level
intuitive ideas on hadron structure.
Various versions of this model have been used for calculating hadronic
matrix elements of weak currents.

Recently, it has become clear that
for a consistent and successful application of quark models to electroweak
decays,
a relativistic treatment of quark spins is necessary \cite{aley},
\cite{neubert}.
However, in the first models by Grinstein, Isgur, Scora and Wise (ISGW)
\cite{gisw}, and Wirbel, Stech, and Bauer (WSB) \cite{wsb} quark spins were not
treated
relativistically. A nonrelativistic approach by GISW is based on a successful
potential model for meson spectrum \cite{gi}. For the calculation of the
electroweak form factors, rescaling of the parameter in the form factor
$q^2$--dependence
is used.
Such an alteration has no strong theoretical deduction. In addition, an
extrapolation from the truly nonrelativistic region $q^2\approx q^2_{max}$,
where the
model is rigorously valid, to a highly relativistic point $q^2=0$ is performed.
The first step to the relativistic treatment of meson decays was done in the
WSB approach.
The quark model calculations are performed only at one point $q^2=0$ using the
Infinite Momentum Frame. For the form factor $q^2$--dependence the authors
postulate a
monopole behavior determined by the nearest vector meson state.
Although the model considers the quark motion
relativistically, the quark spins are again treated in a nonrelativistic
manner.
The WSB approach, as well as the GISW model and its modifications
\cite{aw},\cite{si},
have both the theoretical and experimantal objections, namely:
the models do not reproduce the IW scaling of the form factors in the heavy
quark limit,
and fail to describe the data on the widths and polarizations in the
semileptonic $D\to (K,K^*)l\nu$ decays (Table \ref{table:d2k}).
The answer to these difficulties lies in the correct relativistic consideration
of the spins.
The exact solution to this complicated dynamical problem is not known, but a
simplified
self--consistent relativistic treatment of the quark spins can be performed
within
the light--front formalism \cite{lcqm}.
A description of electroweak properties of pseudoscalar mesons
\cite{jaus}--\cite{card} and
transition form factors at $q^2\le 0$ \cite{jaus} was performed in the
framework of this formalism.
The only difficulty with this approach, is that the applicability of the model
is
restricted by the condition $q^2\le 0$, while
the physical region for hadron decays is
$0\le q^2\le (M_1-M_2)^2$, $M_{1,2}$ being the initial and final hadron mass,
respectively.
So, for obtaining the form factors in the physical region and decay widths and
lepton
distributions, assumptions on the form factor
behavior were necessary. A procedure to remedy this difficulty is proposed
here.

We present a formalism for the relativistic description of the form factors of
hadron decays
within the constituent quark model.
For a direct calculation of the transition form factors at timelike momentum
transfers
we use a dispersion formulation of the light--cone constituent quark model
\cite{lcqm}.
Namely, the amplitudes of hadron interactions considered within the framework
of the light--cone
formalism are represented as dispersion integrals over the hadron mass.
After that, the form factors at $q^2>0$ are derived by performing the analytic
continuation from the region $q^2\le 0$. As a result, the form factors in the
whole kinematic
region $q^2\le (M_1-M_2)^2$ are expressed through the light--cone wave function
of a hadron.
The developed formalism is applied to the analysis of the electroweak
properties of pseudoscalar
mesons.

In the next section we demonstrate the equivalence of the light--cone
constituent
quark model and the dispersion
relation approach \cite{akms}, \cite{ammp}. We present all technical details of
the
description of a pseudoscalar meson within the dispersion
relation approach (leptonic weak decay, two--photon decay, elastic
electromagnetic form factor)
and show the results to be equal to those of the light--cone quark model.

Section 3 considers the transition form factors. Firstly, at spacelike momentum
transfers
the light--cone expression is reformulated as a double dispersion integral
representation.
Secondly, the analytic continuation to the region of timelike
momentum transfers is performed. Along with the normal Landau singularities,
the anomalous non--Landau singularities contribute to the transition form
factors in this region.

In Section 4 the electroweak properties of pseudoscalar mesons are considered.
The following issues are addressed: \\
1. The dependence of the axial--vector decay constant $f_P$ and the
heavy--meson elastic form
factor on the heavy quark mass $m_Q$ is analyzed, using a parameterization of
the meson
wave function based on the heavy quark symmetry. The corrections to the leading
$1/m_Q$--behavior
are estimated to be at the level of $10\div20$\% in the region of $b$-- and
$c$--quark masses. \\
2. The form factors of pseudoscalar meson decays are calculated. Our results
are in agreement with
the QCD sum rules and the experimental data. The form factors can be
approximated to
a high accuracy by the dipole formula in the physical region, but do not
contradict to the
vector meson dominance as well. \\
3. The correlation between the values
of $f_D$ and $f_B$ and the slope of the Isgur--Wise function $\rho^2$ is
studied. The parameter
$\rho^2$ is found to be in the range $0.7\le\rho^2 \le 0.9$ for reasonable
values of
heavy--meson decay constants. We also discuss possible reasons of the deviation
from unity of the
Isgur--Wise function at zero recoil.

The results are summarized in the Conclusion. The appendix provides relevant
technical
details of the dispersion approach.

\newpage
\section{Quark structure of pseudoscalar mesons}
An approach to a composite system description based on dispersion relations
\cite{akms} allows constructing relativistic and
gauge invariant amplitude of the interaction of a composite system
with an external vector field
starting with low-energy constituent scattering amplitude (see the Appendix A).
Two-particle $s$-channel interactions are consistently taken into account both
in the
constituent scattering amplitude and the amplitude of interaction with an
external
field. In the case of a bound state, its
form factor  and structure function are expressed through form factor  and
structure function
of mass-shell constituents and the vertex $G$ of constituent--bound state
transition.
This vertex is defined by the two--particle irreducible block
of the constituent scattering amplitude.
On the one hand, the dispersion integral representation turns out to be
equivalent to the Bethe--Salpeter treatment with a separable kernel of
a special form, the vertex $G$ being
connected with the amputated Bethe-Salpeter wave function of the bound state
\cite{ammp}.
On the other hand, this approach is equivalent to the light--cone description
of a bound state with the special form of spin transformation (the Melosh
rotation).
The vertex $G$ determines the light--cone wave function of the bound state.
Because of the relativistic invariance, the dispersion integral formulation of
the
light--cone approach
approach does not face the problem
of choosing appropriate component of the current for calculating the amplitudes
of the
bound state interaction.

\subsection{The quark--meson vertex}
We discuss the case of a pseudoscalar meson, but the same procedure can be
applied to other hadrons
as well.
The pseudoscalar meson
$P$ with the mass $M$ is considered to be a bound state
of the constituent quark with the mass $m_1$ and the antiquark with the mass
$m_2$.
To derive the expressions for the soft amplitudes of the meson interactions
like
$<\mu\nu|P>$,$<\gamma\gamma|P_0>$, and $<P'|J_\mu|P>$, we start with the
corresponding amplitudes
of the constituent
quark interactions $<\mu\nu|Q\bar Q>$, $<\gamma\gamma|Q\bar Q>$, and $<Q\bar
Q|J_\mu|Q\bar Q>$ and
single out the poles corresponding
to the meson. The amplitude of the $Q\bar Q$ interaction turns out to be the
basic quantity
describing
constituent--quark structure of the bound state. Near a bound state with the
$J^P=0^-$,
the amplitude is dominated by the $S$-wave partial amplitude
which can be expressed in the two--particle approximation through
the dispersion loop graph $B_{ps}$ with the vertex
\begin{equation}
\label{vert}
\frac{\bar Q^a(k_1,m_1) i\gamma_5 Q^a(-k_2,m_2)}{\sqrt{N_c}}G(P^2)
\end{equation}
with
$a$ a color index, $N_c=3$ the number of quark colors,
$k_1^2=m_1^2$, $k_2^2=m_2^2$,
and $P=k_1+k_2$, $P^2\equiv s\ne M^2$.
For on--shell constituents, the expression (\ref{vert}) is the only independent
spinorial structure.

\begin{figure}
\begin{center}
\mbox{\epsfig{file=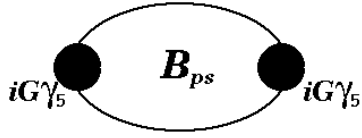,height=2.cm}  }
\end{center}
\caption{Meson dispersion loop graph $B_{ps}(P^2)$.\label{fig:1}}
\end{figure}

The dispersion loop graph Fig.\ref{fig:1}, which is connected
with the meson vertex normalization, reads
\begin{equation}
B_{ps}(P^2)=\int\limits^\infty_{(m_1+m_2)^2}\frac{ds\;G^2(s)}{\pi(s-P^2)}\rho_{ps}(s),
\quad B_{ps}(M^2)=1
\end{equation}

with $\rho_{ps}(s)$ the spectral density of the Feynman loop graph

\begin{equation}
\rho_{ps}(s,m_1,m_2)=-\frac1{8\pi^2}\int dk_1
dk_2\delta(k^2_1-m_1^2)\delta(k^2_2-m_2^2)
\delta(P-k_1-k_2)\;
Sp\left({ (\hat k_1+m_1)i\gamma_5(m_2-\hat k_2)i\gamma_5 }\right)
\end{equation}
$$
=\frac{\lambda^{1/2}(s,m_1^2,m_2^2)}{8\pi
s}(s-(m_1-m_2)^2)\;\theta(s-(m_1+m_2)^2),
$$
where
$$\lambda(s,m_1^2,m_2^2)\equiv(s+m^2_1-m^2_2)^2-4sm_1^2.$$
Taking into account constituent--quark rescatterings leads to the
renormalization of $G$ (see the
Appendix A), and the soft constituent--quark structure of the pion is given by
the vertex

\begin{equation}
\label{vertex}
\frac{\bar Q^a(k_1,m_1) i\gamma_5 Q^a(-k_2,m_2)}{\sqrt{N_c}}G_v(P^2)
\end{equation}
where $G_v(s)=G(s)/B'(M^2)$,
\begin{equation}
\label{vertnorm}
\int \frac{G^2_v(s)\rho_{ps}(s,m_1,m_2)ds}{\pi(s-M^2)^2}=1
\end{equation}
Once the soft vertex (\ref{vertex}) is fixed,
we can proceed with calculating meson interaction amplitudes.

\subsection{Weak decay of a pseudoscalar meson}

Let us consider the decay $P\to\mu\nu$. The corresponding amplitude
reads
\begin{equation}
<P|A^{aa}_\mu(0)|0>=iP_\mu\;f_P,
\end{equation}
$f_P$ is the meson axial--vector decay constant.
To obtain the expression for this matrix element we must first consider the
quantity
\begin{equation}
<Q\bar Q|A^{aa}_\mu(0)|0>
\end{equation}
with $Q$ a constituent quark, while the axial current
$A_\mu(0)=\bar q(0)\gamma_\mu\gamma_5 q(0)$ is defined through current quarks.
Next, we must single out the pole corresponding to the pion.

The $bare$ matrix element has the structure
\begin{equation}
<Q(k_1)\bar Q(k_2)|A^{aa}_\mu(0)|0>_{bare}=\bar Q(k_1)\left[{
\gamma_\mu\gamma_5g^0_A(P^2)+P_\mu\gamma_5h^0_+(P^2)+(k_1-k_2)_\mu\gamma_5h^0_-(P^2)
}\right]Q(-k_2)
\end{equation}
If current quarks were identical to constituent ones we would have had
\begin{equation}
\label{barevalues}
g_A^0(P^2) \equiv 1, \quad h_+^0(P^2) \equiv 0, \quad h_-^0(P^2) \equiv 0.
\end{equation}
It is reasonable to assume that at least at $P^2=(m_1-m_2)^2$ the form factors
$g$ and $h$
are not far from these values.

\begin{figure}
\begin{center}
\mbox{\epsfig{file=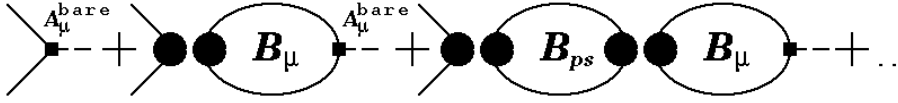,width=12.cm}}
\end{center}
\caption{The series of dispersion graphs for $<Q\bar Q|A_\mu(0)|0>$.
\label{fig:2}}
\end{figure}

The rescatterings of the constituent quarks lead to the series of the
dispersion graphs of
Fig.\ref{fig:2}.
The $bare$ matrix element enters into a single loop graph $B_\mu$
whose spectral density is the product of $G(s)$ and the corresponding Feynman
graph
spectral density which reads
\begin{equation}
\label{fpdensity}
-\frac{\sqrt{N_c}}{8\pi^2}\int dk_1
dk_2\delta(k^2_1-m_1^2)\delta(k^2_2-m_2^2) \delta(P-k_1-k_2)\;
\end{equation}
$$
\times Sp\left({
[\gamma_\mu\gamma_5 g_A^0(P^2)+P_\mu \gamma_5h_+^0(P^2)+(k_1-k_2)_\mu
\gamma_5h_-^0(P^2)]
(\hat k_1+m_1)\;i\gamma_5(m_2-\hat k_2)}\right)
$$
The trace is equal to
\begin{equation}
\label{fptrace}
-4i(k_{1\mu}m_2+k_{2\mu}m_1)g^0_A(P^2)+4iP_\mu(k_1k_2+m_1m_2)h^0_+(P^2)
+4i(k_1-k_2)_\mu(k_1k_2+m_1m_2)h^0_-(P^2)
\end{equation}
So the expression for the loop graph $B_\mu$ takes the form
\begin{equation}
\label{bmu}
B_\mu=4iP_\mu\sqrt{N_c}
\int\limits_{(m_1+m_2)^2}^\infty\frac{ds\;G(s)}{\pi(s-M^2)}\frac{\lambda^{1/2}(s,m_1^2,m_2^2)}{16\pi s}
\frac{s-(m_1-m_2)^2}{2s}
\end{equation}
$$
\times[(m_1+m_2)g_A^0(P^2)-s\;h^0_+(P^2)-(m_1^2-m_2^2)h^0_-(P^2)]
$$

The amplitude with the quark rescatterings taken into account has the same
spinorial
structure as the $bare$ amplitude
\begin{equation}
\label{qqbartomunu}
<Q(k_1)\bar Q(k_2)|A^{aa}_\mu(0)|0>=
\bar Q(k_1)\left[{
\gamma_\mu\gamma_5g_A(P^2)+P_\mu\gamma_5h_+(P^2)+(k_1-k_2)_\mu\gamma_5h_-(P^2)
}\right]Q(-k_2)
\end{equation}
with
$$
g_A(P^2)= g_A^0(P^2)
$$
$$
h_-(P^2)= h_-^0(P^2)
$$
$$
h_+(P^2)= h_+^0(P^2)-\frac{G(P^2)}{1-B_{ps}(P^2)}
$$
$$
\times\int\frac{ds\; G(s)}{\pi(s-M^2)}\frac{\lambda^{1/2}(s,m_1^2,m_2^2)}{16\pi
s}\frac{s-(m_1-m_2)^2}{2s}
4[(m_1+m_2)g_A^0(P^2)-s\;h^0_+(P^2)-(m_1^2-m_2^2)h^0_-(P^2)]
$$
The form factor $h_+$ develops a pole at $P^2=M^2$ as $B_{ps}(M^2)=1$.
Near $P^2=M^2$ the pole dominates the amplitude
\begin{equation}
\label{Ptomunu}
<\bar QQ|A_\mu|0>=<\bar QQ|P>\frac1{M^2-P^2}<P|A_\mu|0>+{\rm regular\; terms}.
\end{equation}
Comparing the pole terms in (\ref{qqbartomunu}) and (\ref{Ptomunu}) and using
the relation
$$
<P|Q\bar Q>=\frac{\bar Q i\gamma_5Q}{\sqrt{N_c}}G_v
$$
one finds
\begin{equation}
<P|A^{aa}_\mu(0)|0>=iP_\mu f_P
\end{equation}
with
\begin{equation}
\label{fp}
f_P=4\sqrt{N_c}\int\frac{ds\;G_v(s)}{\pi(s-M^2)}\frac{\lambda^{1/2}(s,m_1^2,m_2^2)}{16\pi s}
\frac{s-(m_1-m_2)^2}{2s}
\end{equation}
$$
\times[(m_1+m_2)g_A^0(P^2)-s\;h^0_+(P^2)-(m_1^2-m_2^2)h^0_-(P^2)]
$$

Assuming that in reality the values of $g_A^0$ and $h^0$ are not far from the
limit (\ref{barevalues}),
we neglect the terms involving $h^0$ and come to the relation
\begin{equation}
\label{decconst}
f_P=4\sqrt{N_c}(m_1+m_2)g_A^0(M^2)\int\limits_{(m_1+m_2)^2}^\infty
\frac{ds\;G_v(s)}{\pi(s-M^2)}\frac{\lambda^{1/2}(s,m_1^2,m_2^2)}{16\pi s}
\frac{(s-(m_1-m_2)^2)}{2s}
\end{equation}

\subsection{The two-photon decay of the neutral pseudoscalar meson}
We consider the decay of the neutral pseudoscalar meson $P_0$ whose constituent
quark structure is
described by the vertex
\begin{equation}
\frac{\bar Qi\gamma_5 Q}{\sqrt{N_c}}G_v(P^2)
\end{equation}
The rate of the decay $P_0\to2\gamma$ can be written as
\begin{equation}
\Gamma=\frac\pi4\alpha^2M^3g^2_{P\gamma\gamma},\quad
g_{P\gamma\gamma}=G_{P\gamma\gamma}(M^2,0,0),
\end{equation}
where the form factor $G_{P\gamma\gamma}$ is connected with the amplitude
\begin{equation}
<0|J^{em}_{\alpha_2}(q_2)J^{em}_{\alpha_3}(q_3)|P>=2\epsilon_{\alpha_2\alpha_3\beta_2\beta_3}
q_2^{\beta_2}q_3^{\beta_3}G_{P\gamma\gamma}(M^2,q^2_2,q^2_3).
\end{equation}
The electromagnetic current $J^{em}_\mu(0)=\bar q(0)\gamma_\mu q(0)$
is defined through current quarks, whereas the meson structure is described
in terms of the constituent quarks.
So, for calculating the meson amplitude
the constituent quark amplitude of the electromagnetic current is necessary.
The latter is assumed to have the following structure
\begin{equation}
\label{constcurrent}
<Q(k')|\bar q(0)\gamma_\mu q(0)|Q(k)>=\bar Q(k')\gamma_\mu Q(k) f_c(q^2),\quad
q=k'-k.
\end{equation}
The constituent charge form factor $f_c(q^2)$ is normalized such that
$f_c(0)=e_c$,
the constituent charge.
The anomalous magnetic moment of the constituent quark
is neglected in the expression (\ref{constcurrent}),
but it can be included into consideration straightforwardly.

\begin{figure}
\begin{center}
\mbox{\epsfig{file=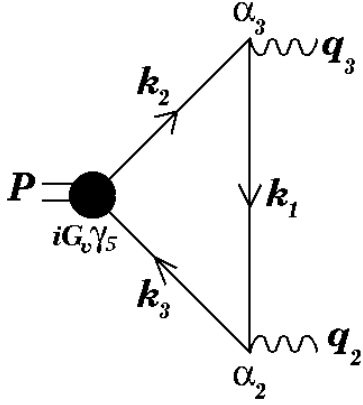,width=5.cm}}
\end{center}
\caption{The graph describing the decay $P^0\to\gamma\gamma$.
\label{fig:p2g}}
\end{figure}

The single dispersion representation for the form factor $G_{P\gamma\gamma}$
reads
\begin{equation}
\label{pg}
G_{P\gamma\gamma}(M^2,s_2,s_3)=f_c(s_2)f_c(s_3)\int\frac{ds_1G_v(s_1)}{\pi(s_1-M^2)}
\Delta_{P\gamma\gamma}(s_1,s_2,s_3),\quad s_2=q^2_2, s_3=q^2_3
\end{equation}
where $\Delta_{P\gamma\gamma}$ is determined by the spectral density of the
Feynman graph of Fig.\ref{fig:p2g}
$$
-\frac{\sqrt{N_c}}{8\pi^2}\int dk_1
dk_2 dk_3 \delta(P-k_2-k_3)\delta(k_2-k_1-q_3)
\frac{\delta(k^2_2-m^2)\delta(k^3_2-m^2)}{m^2-k_1^2}
$$
\begin{equation}
\label{pden}
\times Sp\left({
i\gamma_5(m-\hat k_3)\gamma_{\alpha_2}(m+\hat k_1)\gamma_{\alpha_3}(m+\hat k_2)
}\right)=-\epsilon_{\alpha_2\alpha_3\beta_2\beta_3}q_2^{\beta_2}q_3^{\beta_3}
\Delta_{P\gamma\gamma}(s_1,s_2,s_3)
\end{equation}
The trace reads
\begin{equation}
Sp\left({
i\gamma_5(m-\hat k_3)\gamma_{\alpha_2}(m+\hat k_1)\gamma_{\alpha_3}(m+\hat k_2)
}\right)=4m\epsilon_{\alpha_2\alpha_3\beta_2\beta_3}q_2^{\beta_2}q_3^{\beta_3},
\end{equation}
and we find
\begin{equation}
\label{pgden}
\Delta_{P\gamma\gamma}(s_1,s_2,s_3)=\frac{m\sqrt{N_c}}{4\pi}
\frac{\theta(s_1-4m^2)}{\lambda^{1/2}(s_1,s_2,s_3)}
\log
\left({
\frac{s_1-s_2-s_3+\lambda^{1/2}(s_1,s_2,s_3)\sqrt{1-4m^2/s}}
{s_1-s_2-s_3-\lambda^{1/2}(s_1,s_2,s_3)\sqrt{1-4m^2/s}}
}\right)
\end{equation}
Substituting (\ref{pgden}) into (\ref{pg}), one obtains the expression which
defines the quantity
$G_{P\gamma\gamma}$ for off-shell photons. For real photons one finds
\begin{equation}
g_{P\gamma\gamma}=\frac{m\sqrt{N_c}}{4\pi}e_c^2\int\limits_{4m^2}^{\infty}
\frac{dsG_v(s)}{\pi(s-M^2)}
\frac1s
\log\left({
\frac{1+\sqrt{1-4m^2/s}}{1-\sqrt{1-4m^2/s}}
}\right)
\end{equation}

\subsection{The elastic electromagnetic form factor}

The elastic electromagnetic form factor of a pseudoscalar meson is given by the
following matrix element
\begin{equation}
<P'_M|J^{em}_\mu(0)|P_M>=(P'_M+P_M)_\mu\;F^{el}(q^2)
\end{equation}
$$P_M^2=P'^2_M=M^2,\;P_M-P'_M=q,\;q^2<0.$$
Assuming the following structure for the constituent--quark matrix element of
the electromagnetic
current $J^{em}_\mu(0)=\bar q(0) \gamma_\mu q(0)$,
\begin{equation}
<Q(k'_1)|\bar q(0) \gamma_\mu q(0)|Q(k_1)>=\bar Q(k'_1)\gamma_\mu Q(k_1)
\;f_c(q^2),
\end{equation}
the elastic charge form factor of the meson can be written in the form
\begin{equation}
F^{el}(q^2)=f_1(q^2) H(q^2,m_1^2,m_2^2)+f_2(q^2) H(q^2,m_2^2,m_1^2)
\end{equation}
in terms of the form factors $H$. The quantity $H(q^2,m_1^2,m_2^2)$ describes
the subprocess when
the constituent $1$ interacts with the photon, while the constituent $2$
remains spectator.

\begin{figure}
\begin{center}
\mbox{\epsfig{file=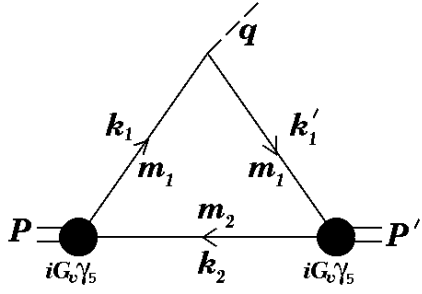,height=4.cm}}
\end{center}
\caption{The contribution $H(q^2,m_1^2,m_2^2)$ to the elastic form factor.
\label{fig:4}}
\end{figure}

The double dispersion representation for the form factor $H(q^2,m_1^2,m_2^2)$
(Fig.\ref{fig:4})
reads
\begin{equation}
\label{ffh}
H(q^2,m_1^2,m_2^2)=
\int\frac{ds\;G_v(s)}{\pi(s-M^2)}
\frac{ds' G_v(s')}{\pi(s'-M^2)}\Delta(s',s,q^2|m_1,m_1,m_2).
\end{equation}
Here $\Delta$ is the double spectral density over $P^2$ and $P'^2$ of the
corresponding triangle
Feynman graph
$$
-\frac1{8\pi}\int dk_1
dk'_1 dk_2 \delta(k^2_1-m_1^2)\delta(k'^2_1-m_1^2)\delta(k^2_2-m_2^2)
\delta(P-k_1-k_2)\delta(P'-k'_1-k_2)
$$
\begin{equation}
\label{deltapi}
\times Sp\left({ (\hat k'_1+m_1)\gamma_\mu(\hat k_1+m_1)
i\gamma_5(m_2-\hat k_2)i\gamma_5 }\right)=2P_\mu(q)
\Delta(s',s,q^2|m_1,m_1,m_2)
\end{equation}
with
$$
P_\mu(q)=(P-\frac{qP}{q^2}\,q)_\mu,\;P^2=s,\;P'^2=s',\;(P'-P)^2=q^2.
$$
The trace reads
\begin{equation}
\label{trace1}
\frac14 Sp\left({ (\hat k'_1+m_1)\gamma_\mu(\hat k_1+m_1)
\gamma_5(m_2-\hat k_2)\gamma_5 }\right)=
2k'_{1\mu} (s-(m_1-m_2)^2)+2k_{1\mu} (s'-(m_1-m_2)^2)+2k_{2\mu}q^2
\end{equation}
Multiplying both sides of (\ref{deltapi}) by $P_\mu$ and using (\ref{trace1})
one obtains at $q^2<0$
\begin{equation}
\Delta(s',s,q^2|m_1,m_1,m_2)=\frac{-q^2}{4\lambda^{3/2}(s',s,q^2)}
\left({
s's+(s'+s-q^2)m_2(m_1-m_2)-(m_1+m_2)(m_1-m_2)^3
}\right)
\end{equation}
$$
\theta(s-(m_1+m_2)^2)\theta(s'-(m_1+m_2)^2)
\theta\left({
-q^2(s'+s-q^2+2(m_1^2-m_2^2))^2
+\lambda(s',s,q^2)(q^2-4m_1^2)
}\right),\quad q^2<0
$$
with $\lambda(s',s,q^2)=(s'+s-q^2)^2-4s's$.

At $q^2=0$ one finds
\begin{equation}
\Delta(s',s,q^2=0|m_1,m_1,m_2)=\pi\rho_{ps}(s,m_1,m_2)\;\delta(s'-s),
\end{equation}
and
\begin{equation}
\label{ffpiat0}
F^{el}(0)=(e_1+e_2)\int\limits_{(m_1+m_2)^2}^\infty\frac{ds\;
G_v^2(s)}{\pi(s-M^2)^2}
\rho_{ps}(s,m_1,m_2)=e_1+e_2
\end{equation}
As we have pointed out in the Appendix A, this is just the Ward identity
consequence.

To reveal the relationship between the dispersion integral (\ref{ffh})
and the light--cone technique, we introduce the light--cone variables
\begin{equation}
\label{lcvariables}
k_-=\frac{1}{\sqrt{2}}(k_0-k_z);\quad
k_+=\frac{1}{\sqrt{2}}(k_0+k_z);\quad
k^2=2k_+k_--k^2_\perp;
\end{equation}
into the integral representation for the form factor  spectral density
(\ref{deltapi}).
We choose the reference frame in which
$$
P_\perp=0,\quad q_+=0,\quad q^2_\perp=-q^2
$$
that is possible at $q^2<0$.
Performing $k_-$ integration
and setting $(\mu=+)$ in both sides of (\ref{trace1}) one finds
\begin{equation}
\label{specden}
\Delta(s',s,q^2|m_1,m_1,m_2)=
\frac1{16\pi}\int \frac{dx d^2k_\perp}{x(1-x)}
\delta\left({s-\frac{m_1^2}{1-x}-\frac{m_2^2}{x}-\frac{k^2_\perp}{x(1-x)}}\right)
\end{equation}
$$
\times\delta\left({s'-\frac{m_1^2}{1-x}-\frac{m_2^2}{x}-\frac{(k_\perp+xq_\perp)^2}{x(1-x)}}\right)
(s'+s-2(m_1-m_2)^2-\frac{x}{1-x}q^2)
$$
Here we denoted $x=k_{2+}/P_+$ and $k_\perp=k_{2\perp}$.

Substituting (\ref{specden}) into (\ref{ffh}) and performing $s$ and $s'$
integrations, one derives
\begin{equation}
\label{ffpilc}
H(q^2_\perp,m_1,m_2)=\int dx d^2k_\perp
\psi(x,k_\perp)\psi(x,k_\perp+xq_\perp)\beta(x,k_\perp,q_\perp)
\end{equation}
where the radial light-cone wave function of a pseudoscalar meson is introduced
\begin{equation}
\label{lcwf}
\psi(x,k_\perp)=\frac{G_v(s)\sqrt{s-(m_1-m_2)^2}}{\pi^{3/2}\sqrt{8}(s-M^2)\sqrt{x(1-x)}},
\qquad s=\frac{m_1^2}{1-x}+\frac{m_2^2}{x}+\frac{k^2_\perp}{x(1-x)}
\end{equation}
$$
\beta=\frac{s-(m_1-m_2)^2+k_\perp q_\perp/(1-x)}
{\sqrt{s-(m_1-m_2)^2}\sqrt{s'-(m_1-m_2)^2}},
\qquad \beta(q_\perp=0)=1
$$
The quantity $\beta$ accounts for the contribution of spins.
It is different from unity  at $q_\perp\ne 0$ because both the spin-nonflip and
spin-flip amplitudes of the interacting quark contribute.
The eq.(\ref{ffpiat0}) is the normalization condition for the soft radial wave
function
\begin{equation}
\int dx\;d^2k_\perp\;|\psi(x,k_\perp)|^2=1.
\end{equation}

In terms of this wave function, the pseudoscalar meson axial--vector decay
constant $f_P$
is represented as
\begin{equation}
\label{fpilc}
f_P=g_A\frac{\sqrt{N_c}}{\sqrt2 \pi^{3/2}}
\int{dx\;d^2k_\perp}\;\psi(x,k_\perp)\frac{m_2(1-x)+m_1x}{\sqrt{s-(m_1-m_2)^2}}
\end{equation}
This expression can be easily deduced by introducing the light--cone variables
into the dispersion
representation (\ref{fpdensity}), making use of (\ref{fptrace}) and examining
the $\mu=+$ component
of the axial current.

Similarly, introducing the light--cone variables into (\ref{pden}) yields the
following expression
for $g_{P\gamma\gamma}$
\begin{equation}
\label{pi2glc}
g_{P\gamma\gamma}=\frac{m\sqrt{N_c}}{\sqrt{2}\pi^{3/2}}\int\frac{dxd^2k_\perp}{\sqrt{x(1-x)}}
\psi(x,k_\perp)\frac x{(m^2+k^2_\perp)\sqrt{s}}
\end{equation}

The same expressions for the form factor, pseudoscalar meson electroweak
constant, and the
two--photon decay constant as
(\ref{ffpilc})--(\ref{pi2glc}) were derived within the light--cone approach in
refs
\cite{jaus}--\cite{card}. \footnote{Our $G_v(s)$ is just equal to $h_0(P)$ of
ref.\cite{jaus}.}

\section{Form factors of meson transitions}
In this section we examine the electroweak transitions of pseudoscalar mesons.
First, we derive the dispersion representations for transition form factors at
$q^2<0$ and
demonstrate them to be equal to those obtained within the light--cone
calculations.
Second, these dispersion representations allow us to perform the analytic
continuation and derive the form
factors of semileptonic decays of pseudoscalar mesons at $q^2>0$ where the
direct application of the
light--cone technique is hampered by the contribution of pair--creation
subprocesses.

\subsection{The pseudoscalar meson transition form factor at $q^2<0$}
\begin{figure}
\begin{center}
\mbox{\epsfig{file=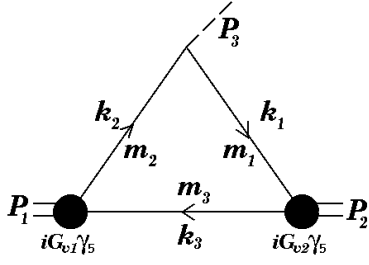,width=5.cm}}
\end{center}
\caption{The dispersion graph for the decay $<P_2|V^{aa}_\mu|P_!>$.
\label{fig:trans12}}
\end{figure}

The amplitude of the weak transition of pseudoscalar mesons $M_1\to M_2$
(Fig.\ref{fig:trans12})
is determined by the two form factors $F_+$ and $F_-$
\begin{eqnarray}
\label{3transamp}
<P_{M_2},M_2|V^{aa}_\mu|P_{M_1},M_1>&=&(P_{M_1}+P_{M_2})F_+(s_3)+(P_{M_1}-P_{M_2})F_-(s_3) \\
<P_{M_2},M_2|A^{aa}_\mu|P_{M_1},M_1>&=&0, \nonumber
\end{eqnarray}
$$
P_{M_2}^2=M_2^2, P_{M_1}^2=M_1^2, P_{M_1}-P_{M_2}=P_{M_3}, P_{M_3}^2=s_3
$$
The weak currents are defined through current quarks
\begin{equation}
\label{3currents}
V^{aa}_\mu=\bar q_1(0) \gamma_\mu q_2(0), \quad A^{aa}_\mu=\bar q_2(0)
\gamma_\mu \gamma_5 q_1(0).
\end{equation}
The structure of the mesons is described in terms of the constituent quarks
by the vertices
\begin{equation}
\label{3vert1}
M_1:\;\frac{\bar Q_2(k_2)i\gamma_5 Q_3(-k_3)}{\sqrt{N_c}}G_{v1}(P_1^2),\quad
M_2:\;\frac{\bar Q_1(k_1)i\gamma_5 Q_3(-k_3)}{\sqrt{N_c}}G_{v2}(P_2^2)
\end{equation}
For calculating the tranition amplitude (\ref{3transamp})
we again need the constituent quark matrix element of the weak current which is
taken in the form
\begin{equation}
\label{3constamp}
<Q_1(k_1)|\bar q_1(0) \gamma_\mu q_2(0)|Q_2(k_2)>=\bar Q_1(k_1)\gamma_\mu
Q_2(k_2)f_{21}(s_3)
\end{equation}
The dispersion representation for the form factors reads
\begin{equation}
\label{3ffs}
F_{\pm}(s_3)=f_{21}(s_3)\int\frac{ds_1G_{v1}(s_1)}{\pi(s_1-M_1^2)}\frac{ds_2G_{v2}(s_2)}{\pi(s_2-M_2^2)}
\Delta_\pm(s_1,s_2,s_3|m_1,m_2,m_3)
\end{equation}
Here $\Delta_\pm$ are the double spectral densities of the Feynman graph
corresponding to
Fig.\ref{fig:trans12} in $s_1-$ and $s_2-$channels
$$
-\frac1{8\pi}\int dk_1
dk_2 dk_3 \delta(k^2_1-m_1^2)\delta(k^2_2-m_2^2)\delta(k^3_2-m_3^2)
\delta(P_1-k_2-k_3)\delta(P_2-k_3-k_1)
$$
\begin{equation}
\label{3delta}
\times Sp\left({ (\hat k_1+m_1)\gamma_\mu(\hat k_2+m_2)
i\gamma_5(m_3-\hat k_3)i\gamma_5 }\right)=(P_1+P_2)_\mu
\Delta_+(s_1,s_2,s_3|m_1,m_2,m_3)
\end{equation}
$$
+(P_1-P_2)_\mu \Delta_-(s_1,s_2,s_3|m_1,m_2,m_3)
$$
with $P_1=P_2+P_3$, $P_1^2=s_1$, $P_2^2=s_2$, $P_3^2=P_{M_3}^2=s_3$.
The vertices $G_{v1,2}$ are normalized in accordance with (\ref{vertnorm}).
The trace reads
\begin{equation}
\label{3trace}
Sp\left({ (\hat k_1+m_1)\gamma_\mu(\hat k_2+m_2)
\gamma_5(m_3-\hat k_3)\gamma_5 }\right)
\end{equation}
$$=
2k_{1\mu} (s_1-(m_2-m_3)^2)+2k_{2\mu}
(s_2-(m_3-m_1)^2)+2k_{3\mu}(s_3-(m_1-m_2)^2)
$$
$$
=(P_1+P_2-2k_3)_\mu\left({a(s_1,m_2,m_3)+a(s_2,m_3,m_1)-a(s_3,m_1,m_2)}\right)
$$
$$
+(P_1+P_2)_\mu a(s_3,m_1,m_2)+(P_1-P_2)_\mu
\left({a(s_2,m_3,m_1)-a(s_1,m_2,m_3)}\right)
$$
where $a(s,\mu_1,\mu_2)=s-(\mu_1-\mu_2)^2$.

Making use of the relation
\begin{equation}
P_1+P_2-2k_3=\frac{b_+(s_1,s_2,s_3)}{\lambda(s_1,s_2,s_3)}(P_1+P_2)+
\frac{b_-(s_1,s_2,s_3)}{\lambda(s_1,s_2,s_3)}(P_1-P_2)
\end{equation}
with
\begin{eqnarray}
b_+(s_1,s_2,s_3)&=&-s_3(s_1+s_2-s_3+m_1^2+m_2^2-2m_3^2)-(m_1^2-m_2^2)(s_1-s_2)
 \\
b_-(s_1,s_2,s_3)&=&(m_1^2-m_2^2)(2s_1+2s_2-s_3)-(s_1-s_2)(s_1+s_2-s_3+m_1^2+m_2^2-2m_3^2),
\end{eqnarray}
we come to the following result for $\Delta_\pm$
\begin{equation}
\Delta_\pm(s_1,s_2,s_3|m_1,m_2,m_3)=\frac{B_\pm(s_1,s_2,s_3)}{\lambda(s_1,s_2,s_3)}\Delta(s_1,s_2,s_3|m_1,m_2,m_3)
\end{equation}
$$
B_+(s_1,s_2,s_3)=b_+(s_1,s_2,s_3)(a(s_1,m_2,m_3)+a(s_2,m_3,m_1)-
a(s_3,m_1,m_2))+a(s_3,m_1,m_2)\lambda(s_1,s_2,s_3)
$$
$$
B_-(s_1,s_2,s_3)=b_-(s_1,s_2,s_3)(a(s_1,m_2,m_3)+a(s_2,m_3,m_1)-a(s_3,m_1,m_2))
$$
$$
+(a(s_2,m_3,m_1)-a(s_1,m_2,m_3))\lambda(s_1,s_2,s_3)
$$
Here $\Delta$, the double spectral density in $s_1$ and $s_2-$channels of the
Feynman triangle graph
$\Gamma(P_1^2,P_2^2,P_3^2)$ with scalar constituents, is introduced
\begin{equation}
\label{3scalar}
\Gamma(P_1^2,P_2^2,P_3^2)=\frac{i}{(2\pi)^4}\int
\frac{dk_1 dk_2 dk_3 \delta(P_1-k_2-k_3)\delta(P_2-k_3-k_1)}
{(m_1^2-k_1^2-i0)(m_2^2-k_2^2-i0)(m_3^2-k_3^2-i0)}
\end{equation}
$$
=\int\frac{ds_1}{\pi(s_1-M_1^2)}\frac{ds_2}{\pi(s_2-M_2^2)}
\Delta(s_1,s_2,s_3|m_1,m_2,m_3).
$$
At $s_3<0$ this spectral density reads
$$
\Delta(s_1,s_2,s_3|m_1,m_2,m_3)=\frac{\theta\left({
-b_+^2(s_1,s_2,s_3)-
\lambda(s_1,s_2,s_3)\lambda(s_3,m_1^2,m_2^2)
}\right)
}{16\lambda^{1/2}(s_1,s_2,s_3)},\quad s_3<0
$$
The solution of this $\theta$-function reads
\begin{equation}
\label{limits}
s_2>(m_1+m_3)^2,\quad s_1^-(s_2,s_3)<s_1<s_1^+(s_2,s_3);
\end{equation}
$$
s_1^\pm(s_2,s_3)=-\frac1{2m_1^2}
$$
$$
\times\left({
s_2s_3-s_2(m_1^2+m_2^2)-s_3(m_1^2+m_3^2)+
(m_1^2-m_2^2)(m_1^2-m_3^2)\pm\lambda^{1/2}(s_2,m_3^2,m_1^2)\lambda^{1/2}(s_3,m_1^2,m_2^2)
}\right)
$$
The final dispersion representation for the form factors at $s_3<0$ takes the
form
\begin{equation}
\label{fftrans}
F_{\pm}(s_3)=f_{21}(s_3)\int\limits^\infty_{(m_1+m_3)^2}\frac{ds_2G_{v2}(s_2)}{\pi(s_2-M_2^2)}
\int\limits^{s_1^+(s_2,s_3)}_{s_1^-(s_2,s_3)}\frac{ds_1G_{v1}(s_1)}{\pi(s_1-M_1^2)}
\frac{B_\pm(s_1,s_2,s_3)}{\lambda^{3/2}(s_1,s_2,s_3)}
\end{equation}
This representation will be the starting point for the consideration of the
meson decays
in the next section.

To demonstrate the equivalence of the dispersion method and the light--cone
approach, we turn back to the
equation (\ref{3delta}) and again make use of the light--cone variables
(\ref{lcvariables}),
choosing the reference frame $P_{M_3+}=0, P_{M_1\perp}=0$.
Setting $\mu=+$ and making use of (\ref{3trace}) gives for $\Delta_+$
\begin{equation}
\label{3deltalc}
\Delta_+(s_1,s_2,s_3|m_1,m_2,m_3)=
\frac1{16\pi}\int \frac{dx_1 dx_2 dx_3 }{x_1 x_2
x_3}d^2k_{3\perp}\delta(x_1-x_2)\delta(1-x_1-x_3)
\end{equation}
$$
\times\delta\left({
s_1-\frac{m_2^2}{x_2}-\frac{m_3^2}{x_3}-\frac{k^2_{3\perp}}{x_2x_3}
}\right)
\delta\left({
s_2-\frac{m_1^2}{x_1}-\frac{m_3^2}{x_3}-\frac{(k_{3\perp}+x_3P_{3\perp})^2}{x_2x_3}
}\right)
$$
(Hereafter $x_i=k_{i+}/P_{+}, P_{1+}=P_{2+}=P_{+}, q_\perp=P_{3\perp},
-q_\perp^2=s_3, x\equiv x_3,
k_\perp \equiv k_{3\perp}$.)
Substituting (\ref{3deltalc}) into (\ref{3ffs}) yields the following expression
for the form factor $F_+$
which gives the main contribution to the semileptonic meson decay rate
\begin{equation}
F_{+}(q_\perp^2)=f_{21}(q_\perp^2)
\int\frac{dxd^2k_\perp}{16\pi^3 x(1-x)}
\frac{ G_{v1}(s_1)}{\pi(s_1-M_1^2)}
\frac{ G_{v2}(s_2)}{\pi(s_2-M_2^2)}
\end{equation}
$$
\times\left({
s_1+s_2-(m_1-m_3)^2-(m_2-m_3)^2+\frac{x}{1-x}(-q_\perp^2-(m_1-m_2)^2)
}\right)
$$
Introducing the radial light--cone wave funcion according to (\ref{lcwf}) leads
to the familiar
light--cone expression (cf.\cite{jaus})
\begin{equation}
\label{fpluslc}
F_{+}(q_\perp^2)=f_{21}(q_\perp^2)\int dx d^2k_\perp
\psi_1(x,k_\perp)\psi_2(x,k_\perp+xq_\perp)\beta_+(x,k_\perp,q_\perp);
\end{equation}
$$
\beta_+=\frac{s_1+s_2-(m_1-m_3)^2-(m_2-m_3)^2+\frac{x}{1-x}(-q_\perp^2-(m_1-m_2)^2)}
{2\sqrt{s_1-(m_2-m_3)^2}\sqrt{s_2-(m_3-m_1)^2}}
$$
$$
=\frac{(m_1x+m_3(1-x))(m_2x+m_3(1-x))+k_\perp(k_\perp+xq_\perp)}
{x(1-x)\sqrt{s_1-(m_2-m_3)^2}\sqrt{s_2-(m_3-m_1)^2}}
$$

\subsection{The transition form factors at $q^2>0$}
For the description of decay processes the form factors in the region
$0<s_3<(M_1-M-2)^2$
are necessary, while the light--cone representation (\ref{fpluslc}) is valid
only at $s_3<0$.
For deriving the form factors at $s_3>0$ the dispersion representation
(\ref{fftrans})
turns out to be a convenient starting point. We write this representation in
the following form
\begin{equation}
\label{4ffs}
F(s_3)=f_{21}(s_3)\int\frac{ds_1G_{v1}(s_1)}{\pi(s_1-M_1^2)}\frac{ds_2G_{v2}(s_2)}{\pi(s_2-M_2^2)}
\frac{B(s_1,s_2,s_3)}{\lambda(s_1,s_2,s_3)}\Delta(s_1,s_2,s_3|m_1,m_2,m_3)
\end{equation}
where $\Delta$ is the double spectral density of the Feynman graph $\Gamma$
with scalar constituents
(\ref{3scalar}). This double dispersion representation defines the analytic
function of $s_3$ both
at negative and positive values provided the proper expression for the spectral
density $\Delta$ is
used. It is important to point out that the functions $G_v(s)$ have no
singularities in
the right--hand side of the complex $s$--plane \cite{akms},
and $B$ and $\lambda$ are polynomials.
So the details of the dispersion integration at $s_3>0$ are determined by the
behavior of the
quantity $\Delta$.

A detailed consideration of the double spectral density $\Delta$ for two
massless constituents
was performed in \cite{sr1}. We extend that consideration to the case of
arbitrary nonzero masses. The same analysis of $\Delta$ for arbitrary masses
was done by Azimov \cite{azimov}.

\begin{figure}
\begin{center}
\mbox{\epsfig{file=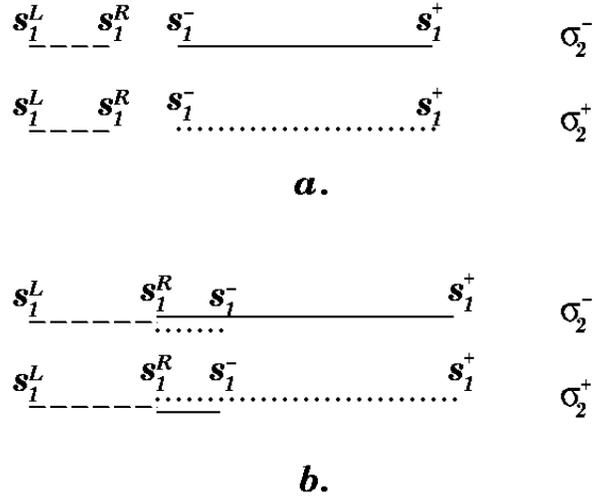,width=8.cm}}
\end{center}
\caption{The location of the singularities of $\sigma_2$ at $s_3>0$: a).
$s_2<s_2^0$; b). $s_2>s_2^0$.
\label{fig:anal}}
\end{figure}

Following \cite{sr1}, we first consider the single dispersion relation in
$P_2^2$.
A standard calculation yields
\begin{equation}
\label{4scalar}
\Gamma(P_1^2,P_2^2,P_3^2)=\int\limits_{(m_1+m_3)^2}^{\infty}
\frac{ds_2}{\pi(s_2-P_2^2)}\sigma_2(P_1^2,s_2,P_3^2),
\end{equation}
where
\begin{equation}
\label{sigma2}
\sigma_2(P_1^2,s_2,P_3^2)=\sigma_+(P_1^2,s_2,P_3^2)-\sigma_-(P_1^2,s_2,P_3^2),
\end{equation}
$$
\sigma_\pm(s_1,s_2,s_3)=\frac1{16\pi\lambda(s_1,s_2,s_3)}
\log\left({
-s_2(s_1+s_3-s_2+m_1^2+m_3^2-2m_2^2)-(s_1-s_3)(m_1^2-m_3^2)
}\right).
$$
Hereafter we assume $m_2>m_1$.
The single dispersion representation reproduces the exact value of the Feynman
expression
(\ref{3scalar}).
Next, we consider the function $\sigma_2(P_1^2,s_2,P_3^2)$ as the analytic
function of
$s_1=P_1^2$ at fixed $s_2$ and $s_3=P_3^2>0$.
As $s_2<s_2^0$ such that
\begin{equation}
\sqrt{s_2^0}=-\frac{s_3+m_1^2-m_2^2}{2\sqrt{s_3}}+
\sqrt{
\left({
\frac{s_3+m_1^2-m_2^2}{2\sqrt{s_3}}
}\right)^2+(m_3^2-m_1^2)
},\quad s_3<(m_2-m_1)^2,
\end{equation}
both of the functions $\sigma_+$ and $\sigma_-$ have square--root branch points
on the physical
sheet at
$s_1^L=(\sqrt{s_2}-\sqrt{s_3})^2$ and $s_1^R=(\sqrt{s_2}+\sqrt{s_3})^2$,
connected by the cut
(dashed line in Fig.\ref{fig:anal}a).

In addition, the function $\sigma_-$ has a logarithmic cut on the physical
sheet from
$s_1^-$ to $s_1^+$ defined by the expression (\ref{limits}).
The square--root cuts cancel in $\sigma_2=\sigma_+ -\sigma_-$, and the
logarithmic cut is the only
singularity of $\sigma_2$ on the physical sheet.
The function $\sigma_+$ has also a
logarithmic cut from $s_1^-$ to $s_1^+$ which is located on the second
unphysical sheet of the
Riemann surface of the square--root (dotted line in Fig.\ref{fig:anal}a),
and does not influence the double spectral density.
The situation changes at $s_2=s_2^0$ which is determined by the condition
$s_1^R(s_2^0)=s_1^-(s_2^0)$. The logarithmic and square--root branch points
coincide, and
for further increasing $s_2>s_2^0$  the logarithm branch point moves up
through the square--root cut onto the physical sheet, whereas the position of
the logarithm
branch point of $\sigma_-$ goes to the second sheet (Fig.\ref{fig:anal}b).
Hence, on the physical sheet the function $\sigma_+$ acquires the logarithmic
cut from
$s_1^-$ to $s_R^-$, and $\sigma_-$ still has the logarithmic cut from $s_1^-$
to $s_+^-$.
Both of the functions have also square--root branch cuts from $s_1^L$ to
$s_+^R$. In the
difference $\sigma_2=\sigma_+-\sigma_-$ the square--root cuts cancel each
other,
but the logarithmic cuts add. The resulting expression for the double spectral
density takes the
form
\begin{equation}
\label{4deltas}
\Delta(s_1,s_2,s_3|m_1,m_2,m_3)=
\frac{\theta(s_2-(m_1+m_3)^2)\theta(s_1^-<s_1<s_1^+)}{16\lambda^{1/2}(s_1,s_2,s_3)}
\end{equation}
$$
+\frac{2\theta(s_3)\theta(s_2-s_2^0)\theta(s_1^R<s_1<s_1^-)}{16\lambda^{1/2}(s_1,s_2,s_3)}
$$
One can check the double dispersion representation (\ref{3scalar}) with the
spectral density
$\Delta$ given by (\ref{4deltas}) to reproduce correctly the Feynman
expression.
The first term in (\ref{4deltas}) relates to the Landau--type contribution
emerging when all
intermediate particles go on mass shell, while the second term describes the
non--Landau
contribution.

In addition to the quantity $\Delta$, the spectral density of the
representation (\ref{4ffs})
involves the factor $1/\lambda(s_1,s_2,s_3)$ which is singular at the lower
limit of
the integration in the non--Landau term, namely
$$
\lambda(s_1,s_2,s_3)=(s-s_1^L)(s-s_1^R).
$$
As it has been discussed in \cite{sr1}, in this case an accurate
application of the Cauchy theorem yields the subtracion term in the non--Landau
contribution.
Representing $\sigma_2$ as a contour integral, we must take into account the
nonvanishing
contribution of the small circle around the point $s_1^R$. Underline once more
that the presence of the factor $G_{v1}(s_1)$ does not change the argumentation
as the function $G_v(s)$ has no singularities at $s_1>(m_2+m_3)^2$.
The final properly regularized representation for the form factors at
$0<s_3<(m_2-m_1)^2$
takes the form (omitting the constituent transition form factor $f_{21}(s_3)$)
\begin{equation}
\label{final}
F(s_3)=
\int\limits_{(m_1+m_3)^2}^\infty\frac{ds_2G_{v2}(s_2)}{\pi(s_2-M_2^2)}
\int\limits_{s_1^-}^{s_1^+}
\frac{ds_1G_{v1}(s_1)}{\pi(s_1-M_1^2)}
\frac{B(s_1,s_2,s_3)}{16\lambda(s_1,s_2,s_3)}
\end{equation}
$$
+
2\theta(s_3)\int\limits_{s_2^0}^\infty\frac{ds_2G_{v2}(s_2)}{\pi(s_2-M_2^2)}
\int\limits_{s_1^R}^{s_1^-}
\frac{ds_1G_{v1}(s_1)}{16\pi(s_1-s_1^R)^{3/2}}
\left[{
\frac{G_{v1}(s_1)B(s_1,s_2,s_3)}{(s_1-s_1^L)^{3/2}(s_1-M_1^2)}-
\frac{G_{v1}(s_1^R)B(s_1^R,s_2,s_3)}{(s_1^R-s_1^L)^{3/2}(s_1^R-M_1^2)}-
}\right]
$$
It should be pointed out, that although the representations (\ref{4ffs}) and
(\ref{final})
were deduced for the case of pseudoscalar mesons, transition form factors of
any hadrons have the
same structure. A particular choise of the initial and final hadrons yields a
specific polynomial
$B$. So the performed analysis is valid in the general case of hadron decay.

\section{Calculation results}
We are now in a position to apply the developed formalism to
the analysis of the properties of pseudoscalar mesons and to the
direct calculation of the decay form factors. To this end
we must specify the parameters of the model, i.e.
input the vertex functions of the pseudoscalar mesons and constituent quark
masses.

\subsection{Parameters of the model}
For a pseudoscalar meson built up of quarks with the masses $m_Q$ and $m_q$,
it is convenient to introduce the function
$\phi$ related to the vertex function $G_v$ as
\begin{equation}
\label{5vertex}
G_v(s)=\frac{\pi}{\sqrt{2}}\frac{\sqrt{s^2-(m_Q^2-m_q^2)^2}}{\sqrt{s-(m_Q-m_q)^2}}\frac{s-M^2}{s^{3/4}}\phi(k),\qquad
k=\frac{\lambda^{1/2}(s,m_Q^2,m_q^2)}{2\sqrt{s}}
\end{equation}
The normalization condition (\ref{vertnorm}) for $G_v$ yields the following
normalization
condition for $\phi$
\begin{equation}
\int\phi^2(k)k^2dk=1.
\end{equation}
The function $\phi$ is the ground--state $S$--wave radial wave function of a
pseudoscalar meson
for which we choose a simple exponential form
\begin{equation}
\label{5exppar}
\phi(k)=\exp\left({-4\alpha \frac{k^2}{\mu^2_P}}\right)
\end{equation}
where $\mu_P=m_Qm_q/(m_Q+m_q)$ is the reduced mass.
The parameterization (\ref{5exppar}) is inspired by the nonrelativistic quantum
mechanics and, as we shall see later, is convenient for the analysis of the
case
$m_Q\to\infty$.

In the nonrelativistic quantum mechanics
a bound--state wave function is determined by the motion of the particle
with the mass $\mu_P$ in the potential independent of masses, and thus $\alpha$
does not depend on the masses as well. Relativistic effects
destroy this simple feature of the wave function. In QCD the situation is much
more
complicated because additional dimensional quantities such as $\Lambda_{QCD}$
and the
condensates appear.
So, $\alpha$ should be considered as some unknown function of the quark masses.
It is possible to obtain the information on the behavior of $\alpha$ as a
function of $m_Q$
at fixed $m_q=m_{u,d}=0.25\;GeV$ in the two regions: at small $m_Q$ and
$m_Q\to\infty$.

\begin{figure}
\begin{center} \mbox{\epsfig{file=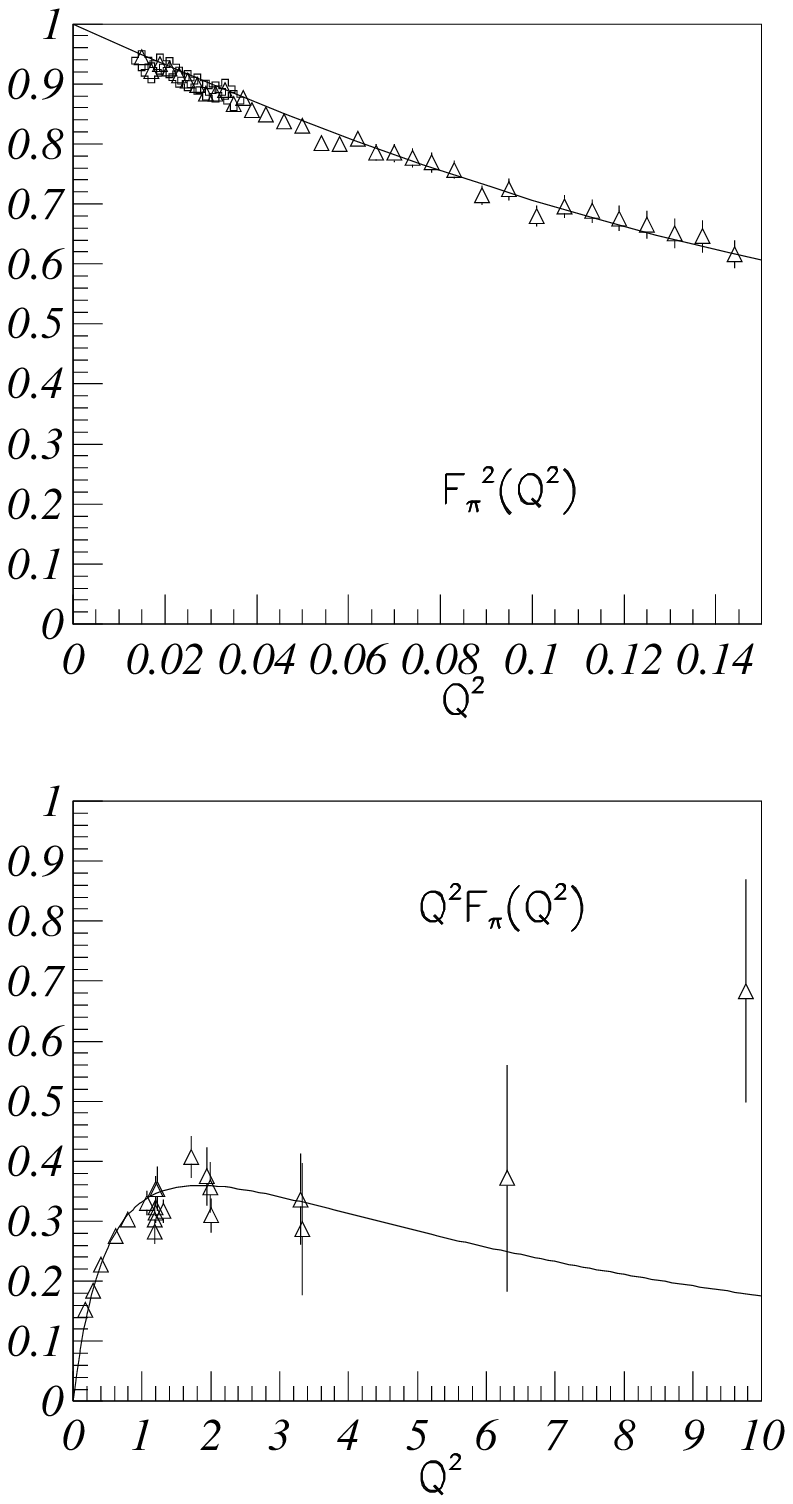,width=12cm}}
\end{center}
\caption{The $\pi^+$ form factor, $\alpha_\pi=0.02$. \label{fig:ffpi}}
\end{figure}

\begin{figure}
\begin{center} \mbox{\epsfig{file=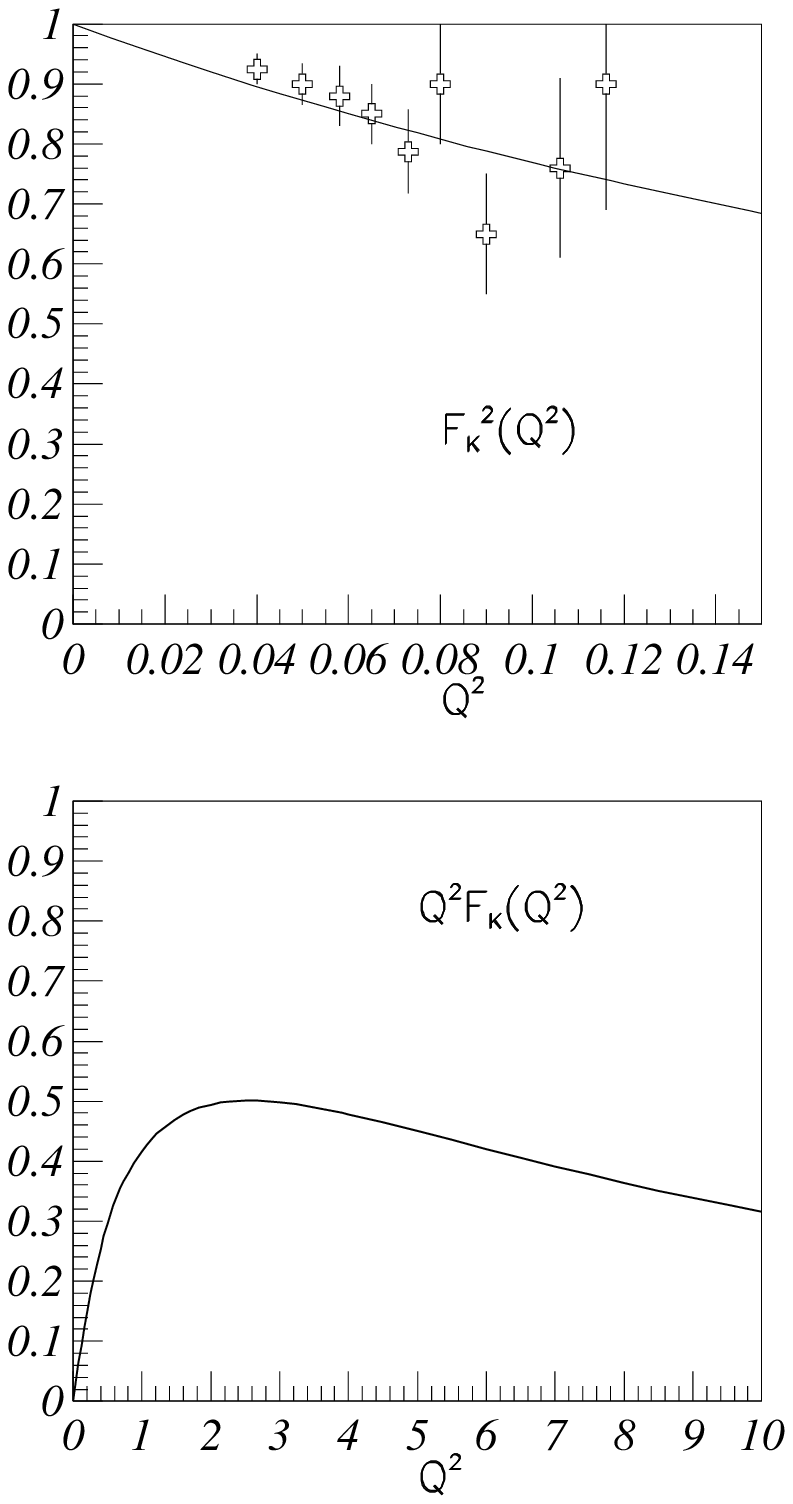,width=12cm}}
\end{center}
\caption{The $K^+$ form factor, $\alpha_K=0.02$. \label{fig:ffk}}
\end{figure}

At $m_Q\le 0.5\;GeV$  the value of $\alpha$ can be determined by
describing the data in the light--meson sector.
The light--quark masses given in Table \ref{table:parameters} and
$\alpha_\pi=\alpha_K=0.02$ provide a good description of the data on
$f_\pi$, $f_K$, and the elastic form factors (Figs. \ref{fig:ffpi} and
\ref{fig:ffk}).
The meson decay constants and form factors are calculated with the values
$g_A^0(M^2)=1$ and $f_c(q^2)=f_c(0)$, respectively.

In the region $m_Q\to\infty$ the behavior of $\alpha$ can be found on the basis
of the heavy quark symmetry.
To this end, let us consider the amplitudes of the elastic and inelstic
transitions between
pseudoscalar mesons consisting of heavy $Q$ and light $q$ quarks and
introduce the dimensionless form factors as follows
\begin{equation}
\label{5el}
<M,P'|\bar Q\gamma_\mu Q|M,P>=(P'+P)\;F_{el}(q^2);\quad q^2\le 0
\end{equation}
$$
F_{el}(q^2)=h_{el}(\omega)=1-\rho^2_{el}(\omega-1)+O((\omega-1)^2),\quad
\omega=1-\frac{q^2}{2M^2}\ge 1
$$
\begin{equation}
\label{5inel}
<M_{2},P_2|\bar Q_2\gamma_\mu
Q_1|M_{1},P_1>=(P_1+P_2)F_{+}(q^2)+(P_1-P_2)F_{-}(q^2); \qquad
0<q^2<(M_1-M_2)^2
\end{equation}
$$
h_{\pm}(\omega)=\frac{M_1\pm M_2}{2\sqrt{M_1M_2}}F_{+}(q^2)+\frac{M_1\mp
M_2}{2\sqrt{M_1M_2}}F_{-}(q^2);\quad
\omega=\frac{M_1^2+M_2^2-q^2}{2M_1M_2}\ge 1.
$$
$$
h_{+}(\omega)=h_+(1)-\rho^2(\omega-1)+O((\omega-1)^2).
$$
In the limit of infinitely heavy quarks $Q_{1,2}$,
the amplitudes are expressed in terms of the single universal Isgur--Wise
function (IW) $\xi(\omega$) \cite{iw}
\begin{equation}
\label{5hqlimit}
h_+(\omega)=h_{el}(\omega)=\xi(\omega), \quad h_-(\omega)=0, \quad
\xi(\omega)=1-\rho^2(\omega-1)+O((\omega-1)^2).
\end{equation}
In addition, the qeavy quark symmetry predicts the universal relation for
heavy--meson decay constants
\begin{equation}
\label{5fp}
\sqrt{M_P}f_Q=const.
\end{equation}
The asymptotic relations (\ref{5hqlimit}) and (\ref{5fp}) are the zero--order
terms of
the $1/m_Q$--expansion which is calculable within the
HQET \cite{hqet}. A particular form of the IW function depends on the heavy
meson wave function.

The expressions (\ref{5hqlimit}) and (\ref{5fp}) mean that the HQ symmetry
restricts
the possible behavior of the meson wave function at large $m_Q$.
Table \ref{table:heavymeson} gives the results on
$f_P$ and $\rho^2_{el}$ vs $m_Q$ at $m_q=0.25\;GeV$, and
Fig.\ref{fig:fp} presents the quantity $\sqrt{m_Q}f_P$ as the function of
$m_Q$ for various values of $\alpha$.
In the HQ limit, for a finite binding energy of the meson
the heavy meson and the heavy quark masses coincide, $M_Q/m_Q=1$.
So, the value of $\sqrt{m_Q}f_P$ should be independent of the heavy quark mass.

These results show that
the asymptotic relations (\ref{5hqlimit}) and (\ref{5fp}) are satisfied
if the parameter $\alpha$ of the wave function (\ref{5exppar})
tends to a constant $\alpha_\infty$ as $m_Q\to\infty$.

Thus, the function $\alpha(m_Q)$ has the following behavior: it is equal to
0.02 at $m_Q\le0.5\;GeV$ and
tends to a constant $\alpha_\infty$ as $m_Q\to \infty$. For investigating the
$B$ and $D$ mesons and
their decays we need the information on $\alpha$ in the region $m_Q=2\div
5\;GeV$.

The simplest way is to extract $\alpha$ at $m_Q=2\div5\;GeV$ from the
analysis of $f_D$ and $f_B$ as we have done for the light mesons.
In the absence of the experimental data we refer to the results of other
models.
As one can see, the decay constants $f_P$ calculated with $\alpha$ from
the range $0.02\le\alpha_D,\alpha_B\le 0.04$ cover the regions
$160\;MeV\le f_D \le 230\;MeV$ and $130\;MeV\le f_B \le 200\;MeV$ which include
the
predictions of most of the models.
Hence, the values of $\alpha_D$ and $\alpha_B$
related to the true wave functions of $D$ and $B$ mesons
are expected to be inside the interval $0.02\div 0.04$.

However, there is an attractive possibility to specify $\alpha_{D,B}$ more
precisely.
Namely, it seems reasonable to assume $\alpha$ to be
approximately constant in the region $m_Q\ge 1\div 2\;GeV$.
There are at least two arguments behind this assumption.
Firstly, a system consisting of a heavy and a
light particles behaves like a quasinonrelativistic system. And secondly, there
are no visible sources
within QCD to yield steep changes of $\alpha$ in this region.
Then for the $B$ and $D$ mesons one expects $\alpha_D=\alpha_B=\alpha_\infty$.
The next step is to estimate $\alpha_\infty$.
We consider the value $\alpha_\infty=0.02$ to be both
attractive and reasonable: on the one hand, the same parameter describes all
ground--state mesons, and on the other hand, one finds for $\alpha_\infty=0.02$
$$
\sqrt{m_\infty}f_{P\infty}\simeq 5.8\;GeV^{3/2}
$$
in agreement with the value $0.6\div 0.7$ estimated in \cite{sr5}.

Assuming $\alpha_D=\alpha_B=\alpha_\infty$, we can estimate the magnitude of
the
higher order $1/m_Q$ corrections which determine
the deviations of the calculated $f_P$ and $\rho^2_{el}$ at finite $m_Q$
from the asymptotic relations (\ref{5hqlimit}) and (\ref{5fp}).
Rather strong violation of the HQ symmetry for $b-$ and
$c-$quarks ($5\div15$\% at $m_Q=5\;GeV$ and $20\div30$\% at $m_Q=2\;GeV$) can
be
observed both in $f_P$ and $\rho^2_{el}$ at $\alpha_\infty=0.02\div 0.04$.

We shall analyze the transition form factors obtained at $\alpha_{D,B}=0.02$
and $0.04$.
If our assumption $\alpha_D=\alpha_B=\alpha_\infty=0.02$ does not work
properly,
the form factor calculations for $\alpha=0.02$
and $\alpha=0.04$ give an interval which is expected to include the true value.

Table \ref{table:parameters} gives the numerical parameters of the model.

\subsection{Discussion}
\begin{figure}
\begin{center}
\mbox{\epsfig{file=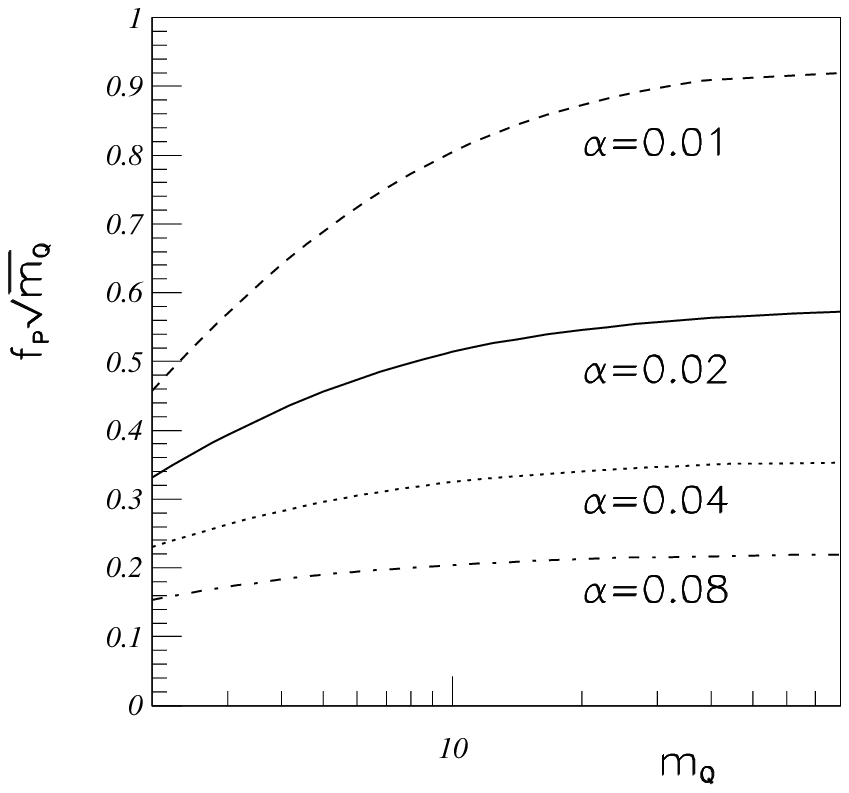,height=10.cm}}
\end{center}
\caption{The quantity $m_Q^{0.5}f_P$ as the function of $m_Q$ at $m_q=0.25\;
GeV$.
\label{fig:fp}}
\end{figure}

\begin{figure}
\begin{center} \mbox{\epsfig{file=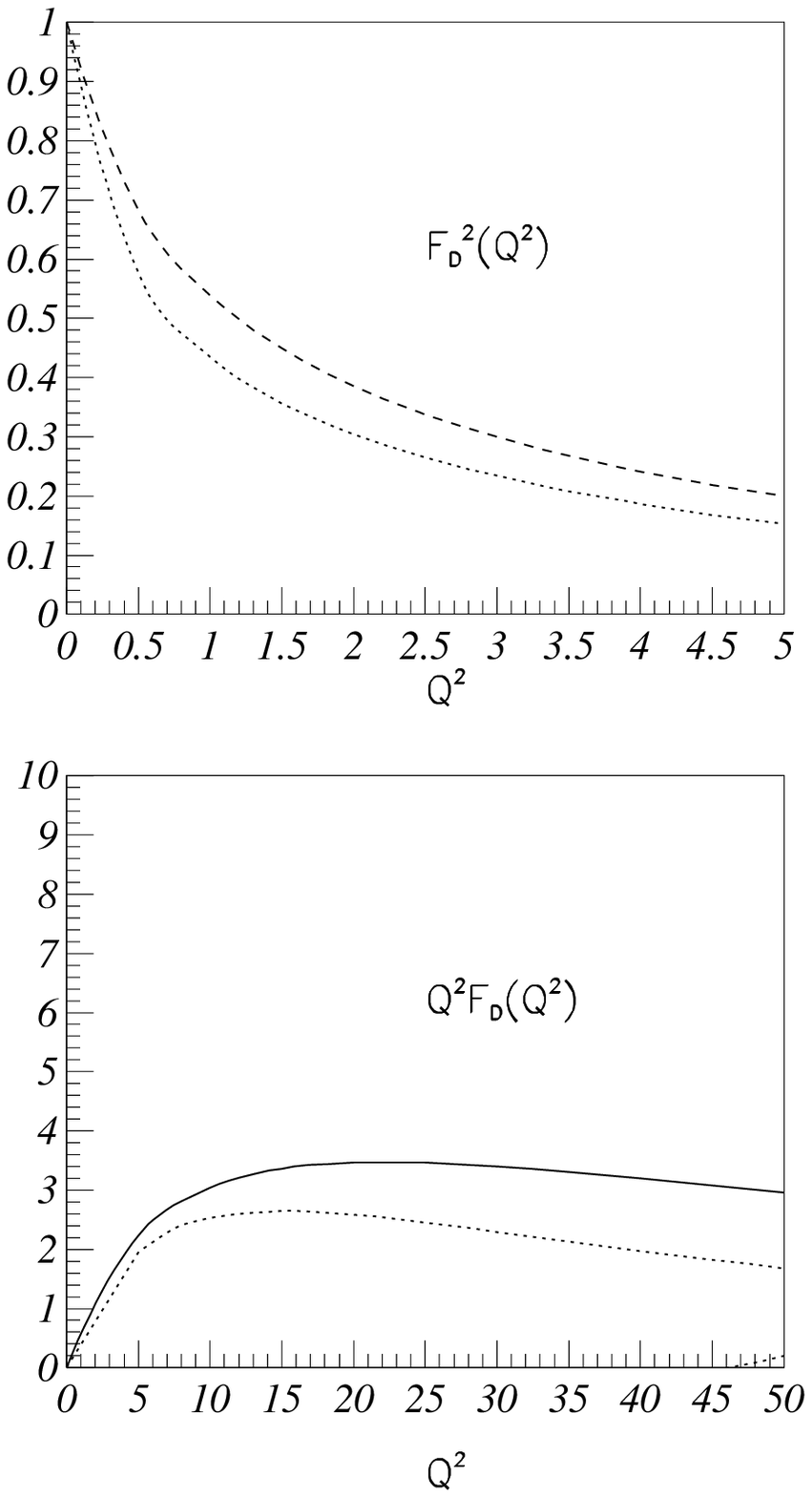,width=12cm}}
\end{center}
\caption{The $D^+$ form factor. Solid -- $\alpha_D=0.02$, dotted --
$\alpha_D=0.04$.
\label{fig:ffd}}
\end{figure}

\begin{figure}
\begin{center} \mbox{\epsfig{file=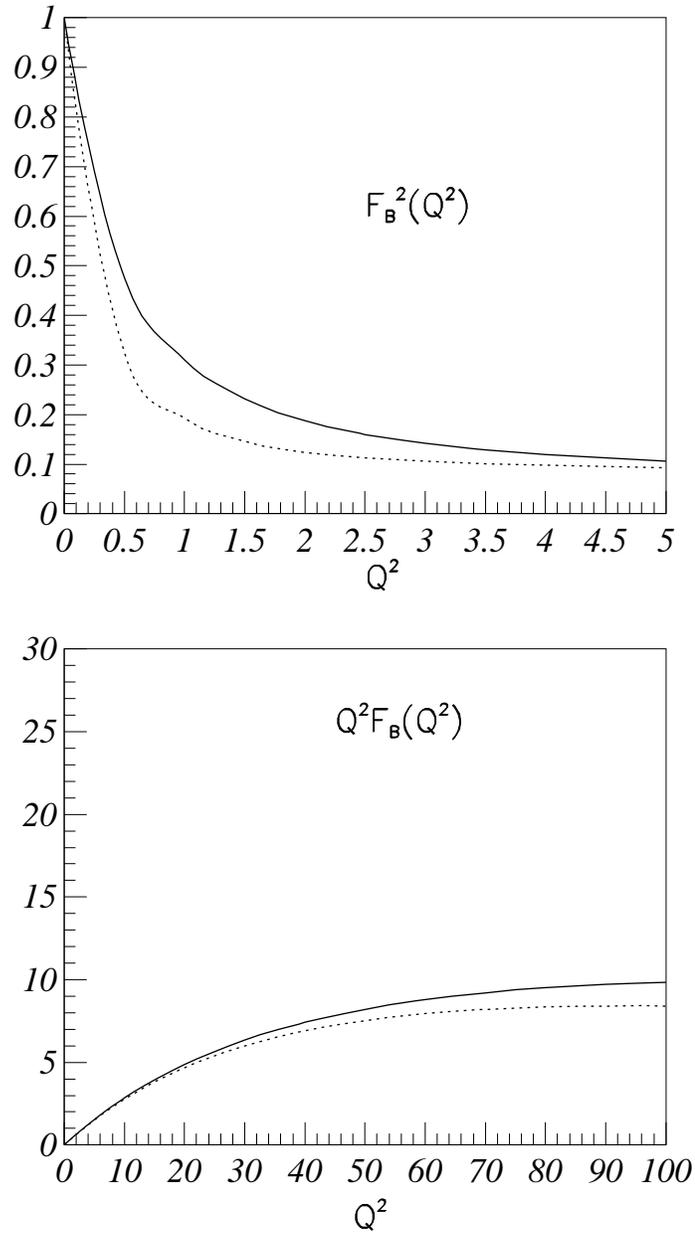,width=12cm}}
\end{center}
\caption{The $B^+$ form factor. Solid -- $\alpha_B=0.02$, dotted --
$\alpha_B=0.04$.
\label{fig:ffb}}
\end{figure}

\begin{figure}
\begin{center} \mbox{\epsfig{file=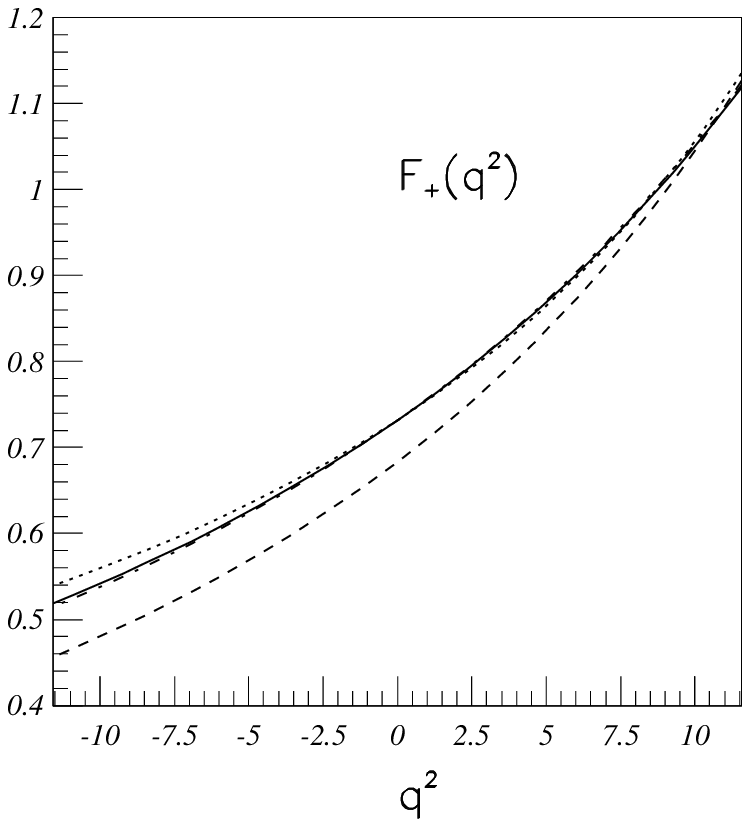,height=10cm}}
\end{center}
\caption{The form factor $F_+(q^2)$ for $B\to D$. Solid --
$\alpha_D=\alpha_B=0.02$,
dotted - the monopole fit, dash--dotted -- the dipole fit. Dashed --
$\alpha_D=\alpha_B=0.04$
\label{fig:b2d}}
\end{figure}

\begin{figure}
\begin{center} \mbox{\epsfig{file=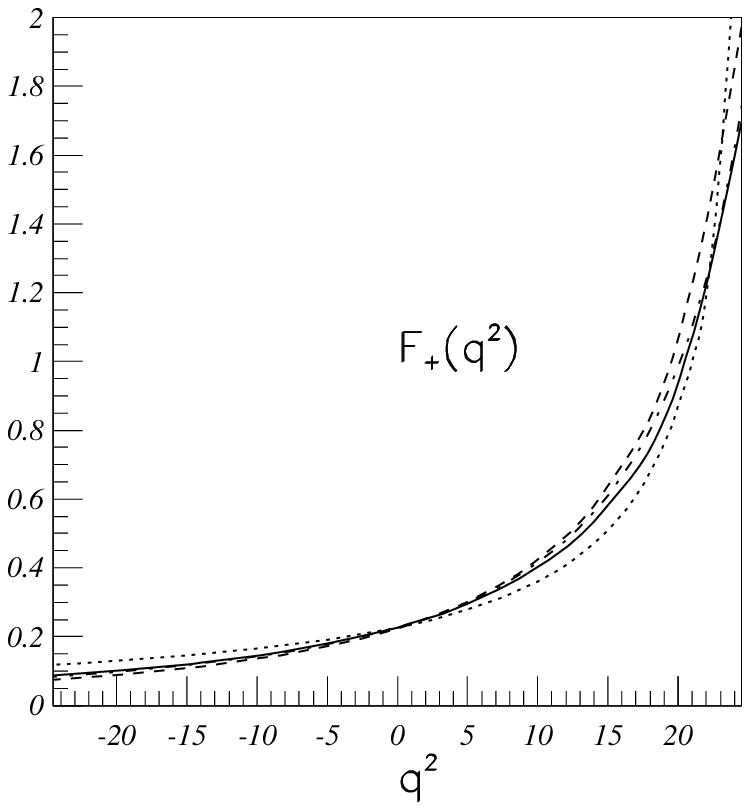,height=10cm}}
\end{center}
\caption{The form factor $F_+(q^2)$ for $B\to \pi$. Solid -- $\alpha_B=0.02$,
dotted - the monopole fit, dash--dotted -- the dipole fit. Dashed --
$\alpha_B=0.04$.
\label{fig:b2pi}}
\end{figure}

\begin{figure}
\begin{center} \mbox{\epsfig{file=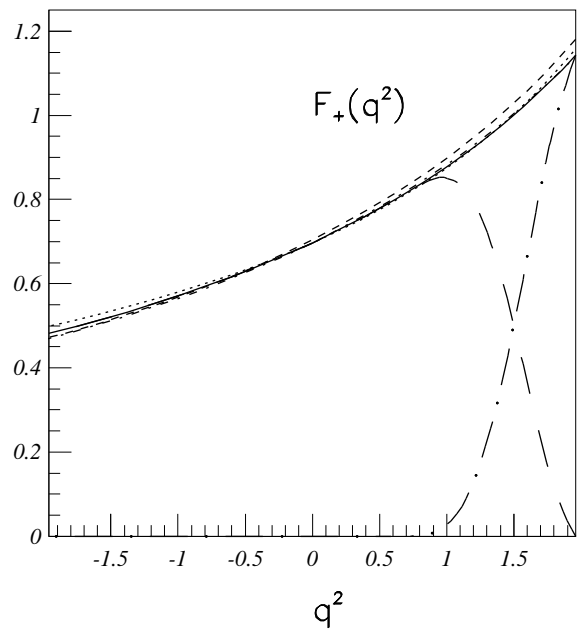,height=10cm}}
\end{center}
\caption{The form factor $F_+$ for $D\to K$. Solid -- $\alpha_D=0.02$,
dotted - the monopole fit, dash--dotted -- the dipole fit.
Long--dashed -- the Landau singularity contribution, long--dash--dotted -- the
non-Landau term.
Dashed -- $\alpha_D=0.04$.
\label{fig:d2k}}
\end{figure}

\begin{figure}
\begin{center} \mbox{\epsfig{file=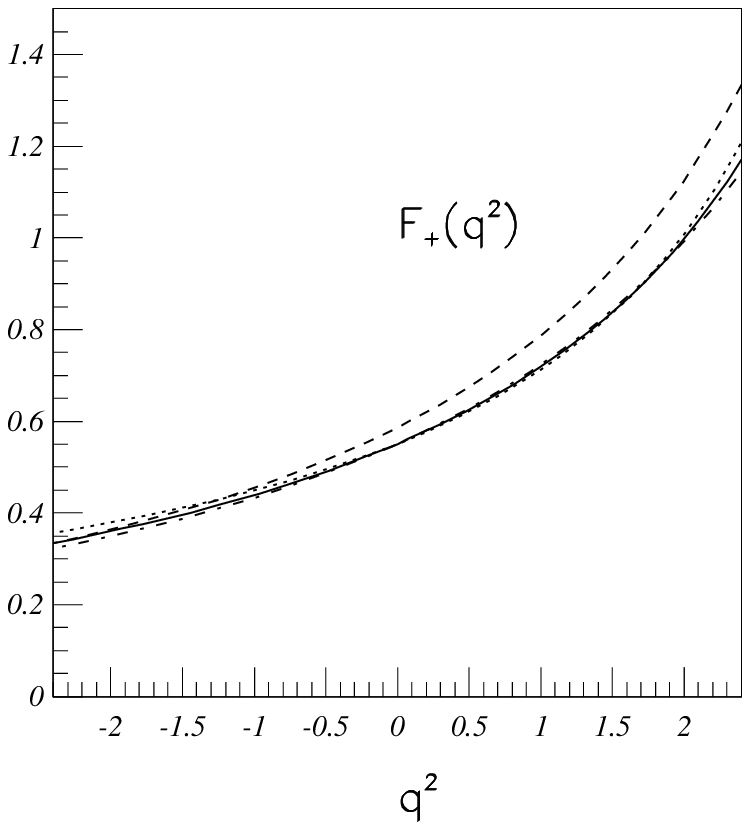,height=10cm}}
\end{center}
\caption{The form factor $F_+$ for $D\to\pi$. Solid -- $\alpha_D=0.02$,
dotted - the monopole fit, dash--dotted -- the dipole fit. Dashed --
$\alpha_D=0.04$.
\label{fig:d2pi}}
\end{figure}

1. The results on the axial--vector decay constant $f_P$ are shown
in Fig.\ref{fig:fp} and Table \ref{table:heavymeson}.
Assuming $\alpha(m_Q)=\alpha_\infty$ at $m_Q\ge2\;GeV$,
one can see the asymptotic relation $\sqrt{m_Q}f_P=const$ to work
perfectly at $m_Q>40-50\;GeV$,
and finds essential corrections to the asymptotic relations at lower $m_Q$.
For $\alpha_\infty=0.02$ one obtains $f_D=234\;MeV$ and $f_B=202\;MeV$ that
confirms the
expectation $f_D\simeq f_B$ \cite{sr4}. These values for the decay constants
correspond to the
constituent quark decay constant $g_A^0=1$. In reality, the latter can be less
than
unity,
$g_A^0\simeq 0.75\div 1$.
This will lead to decreasing the $f_P$. \\
2. Figures \ref{fig:ffd}--\ref{fig:d2pi} present the elastic and transition
form factors calculated with $\alpha_{D,B}=0.02$ and $0.04$.

The $K\to\pi$ transition form factor is well approximate by the linear function
$F_+(q^2)=F_+(0)+aq^2$, $F_+(0)=0.96$, $a=1.27\;GeV^{-2}$ in agreement with the
results of \cite{lr}.

\begin{figure}
\begin{center} \mbox{\epsfig{file=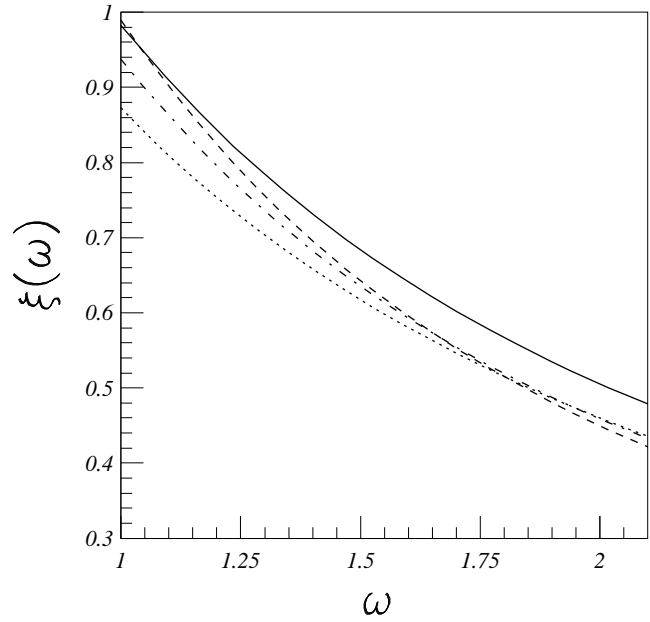,height=10cm}}
\end{center}
\caption{The Isgur--Wise function $\xi(\omega)=h_+(\omega)$ of the decay $B\to
D$:
Solid -- $\alpha_D=\alpha_B=0.02$,
dashed -- $\alpha_D=\alpha_B=0.04$,
dash-dotted -- $\alpha_D=0.04,\;\alpha_B=0.02$,
dotted -- $\alpha_D=0.02,\;\alpha_B=0.04$.
\label{fig:iw}}
\end{figure}

The parameters of the monopole $F_+(q^2)=F_+(0)/(1-q^2/M^2_{mon})$
and the dipole $F_+(q^2)=F_+(0)/(1-q^2/M^2_{dip})^2$ fits to the other
transition form factors
are given in Table \ref{table:fits}.
The dipole formula excellently approximates the
transition form factors with better than 1\% accuracy.
Although the monopole fit provides a worse accuracy, its parameters agree with
the
vector meson dominance.
The values $F_+(0)$ are close to the corresponding results of QCD sum rules
(cf. Tables \ref{table:b2pi} and \ref{table:d2k}) and the existing experimental
data.
\\
3. Fig.\ref{fig:iw} plots the IW function $\xi(\omega)=h_+(\omega)$
for the decay $B\to D$ at various values of $\alpha_D$ and $\alpha_B$.
Table \ref{table:iw} gives the parameters of the calculated IW function.
The function $h_-(\omega)$ turned out to be negligibly
small in agreement with (\ref{5hqlimit}).

The IW function has been extensively studied
both theoretically and experimentally (see Table \ref{table:rho2}). The
analysis by ARGUS
\cite{iwexp1} and most of the earlier theoretical results suggested
$1<\rho^2<2$. However,
a recent analysis by CLEO as well as recent theoretical estimates favor the
lower
values $\rho^2\le1$.
We found the relation $0.7\le\rho^2\le0.9$ for all values of $\alpha_{D,B}$
from the considered
interval.

As it follows from the HQ symmetry, the value $\xi(1)$ strongly depends on the
relationship
between $\alpha_D$ and $\alpha_B$:
it turns out to be close to unity for $\alpha_B=\alpha_D$ and steeply decreases
as $\alpha_B\ne\alpha_D$. One can find rather uncertain constraint for the
considered
region of the parameters $\alpha_{D,B}$
$0.87\le\xi(1)\le 0.98$.

Let us underline that except for the relationship between $\alpha_D$ and
$\alpha_B$,
the value $\xi(1)$ is also affected by the particular values of heavy--meson
binding
energies.
At large quark masses and the binding energy kept finite, the heavy meson and
heavy quark masses coincide, $M_\infty/m_\infty=1$.
Hence, the positions of the 'quark zero recoil point'
$q_0^2=(m_{Q_1}-m_{Q_2})^2$ and the meson zero recoil point
$q_{max}^2=(M_{1}-M_{2})^2$
also coincide. For infinitely heavy quarks this yields $\xi(1)=1$.
For the physical heavy quarks and mesons, the positions of the 'quark zero
recoil point' $\omega_0=1-((m_{Q1}-m_{Q2})^2-(M_1-M_2)^2)/2M_1M_2$ and the
'meson zero recoil point'
$\omega=1$ do not coincide any longer.
The calculated $\xi(\omega_0)$
turns out to be not far from unity if $\alpha_D=\alpha_B$. So, the value
$\xi(1)$ is sensitive to the particular values of the quark masses.
The quark masses used in our calculation are chosen such that
$m_b-m_c=M_B-M_D$,
and thus $q^2_{0}=q^2_{max}$ and $\omega_0=1$.
That is why $\xi(1)\simeq 1$ at $\alpha_D=\alpha_B$.
For other reasonable values of quark masses, the
deviation from unity at $\alpha_D=\alpha_B$ are found at the level of 3--4\%.
\\
4. The analysis of the analytic properties of the hadron transition form
factors
yields the following typical picture demonstrated in Fig.\ref{fig:d2k}:
at $q^2\le 0$ the contribution of the non--Landau singularity is absent, and
the Landau--type
singularity determines the form factor;
in the region $0<q^2<(m_2-m_1)^2$
both of them are essential; at the point $q^2=(m_2-m_1)^2$ the
contribution of the Landau singularity vanishes, and the non--Landau
singularity
determines the decay form factor at this 'quark zero recoil' point.

For hadron decays related to the heavy--to--heavy quark transitions,
a specific relationship between the Landau and the
non--Landau contributions to the dispersion representation is observed:  the
normal Landau contribution dominates the form factor at all $q^2< (m_2-m_1)^2$,
whereas the anomalous singularity is essential only in the close vicinity of
this
point. So, effectively the transition form factor are determined by the
contribution
of the Landau contribution only.
Thus, the HQ symmetry can be formulated in the language of
the analytic properties of the transition form factors as the
dominance of the Landau singularity in the almost whole kinematical region.

In the case of the meson decay related to a heavy--to--light
quark transition, the anomalous non--Landau contribution
is important in a broad kinematical region. So the relations suggested by the
HQ
symmetry would not work properly.

\section{Conclusion}
We investigated form factors of hadron transitions within the relativistic
constituent quark model and proposed a formalism for a direct calculation of
hadron decay
form factors. The developed approach was applied to the analysis of the
electroweak
properties and transitions of pseudoscalar mesons.
Our main results are:\\
1. The equivalence of the light--cone constituent quark model and the
approach based on the dispersion relation integration over a bound state mass
for the description of leptonic decays and transition form factors
at spacelike momentum transfers has been demonstrated.
Although the comparison has been performed for a particular case of
pseudoscalar mesons,
the approaches are equivalent for the description of any hadrons. \\
2. The obtained dispersion formulation of the light--cone constituent quark
model
allows a consideration of the decay processes where the direct application of
the light--cone
technique is hampered by the contribution of pair--creation subprocesses.
The analytic continuation in the dispersion representation of the transition
form factor
yields the form factor at timelike momentum transfers expressed through the
meson radial
light--cone wave function.
Along with the normal Landau singularities, the anomalous non--Landau
singularities contribute
to the form factors at $q^2>0$. \\
3. For hadron decays related to the heavy--to--heavy quark transitions a
specific relationship
between the contributions of the Landau--type and the non--Landau singularities
has been observed.
This allows a formulation of the heavy quark symmetry in the language of the
analytic properties
of the decay
form factors as the dominance of the normal Landau contribution in the almost
whole kinematic
region of momentum transfers. \\
4. Electroweak properties and form factors of pseudoscalar mesons
have been analyzed using a parameterization of the meson wave function based on
the
heavy quark symmetry. We have examined the dependence of the axial--vector
decay constant on the
heavy--quark mass, and found $f_D\simeq 235\;MeV$ and $f_B\simeq200\;MeV$.
These values can be
decreased by a factor of $0.75\div 1$, if the decay constant at the level of
the constituen quarks
is less than unity.

The correlation between the axial--vector decay constant $f_P$ and the
transition form factors
yields the IW function parameter $\rho^2= 0.8\pm0.1$ for the
axial--vector decay constants from the intervals
$160\;MeV\le f_D\le 235\;MeV$ and $130\;MeV\le f_D\le 200\;MeV$

Analyzing the dependence of $f_P$ and the heavy meson form factor on the heavy
quark mass we have found that the violation of the HQ symmetry relations can be
expected at the 10--20\% level for the $b$-- and $c$--quark masses .\\
5. The calculated form factors of pseudoscalar meson transitions have been
approximated with a
1\%--accuracy by the dipole formula in the whole kinematic region.
The form factors are also compatible with the vector meson dominance and are
close to
the results of the QCD sum rules.

The developed approach can be applied to the description of the
pseudoscalar--to--vector
meson transitions and rare decays of heavy mesons. This work is now in
progress.

I am grateful to V.V.Anisovich, Ya.I.Azimov, and K.A.Ter--Martirosyan for
discussing
the general problems and technical details related to hadron decays.
I am also indebted to the German Ministry of Science and Technology
for the financial support at the early stage of this work and to H.R.Petry for
his hospitality
during my stay in Bonn.

\newpage
\begin{table}[1]
\caption{\label{table:b2pi}
The form factors of the decays $B\to \pi,\rho$ at $q^2=0$. The labels Lat, SR,
QM and LCQM stand for Lattice, Sum Rules,
Quark Model and Light Cone Quark Model, respectively. }
\centering
\begin{tabular}{|c|c|c|c|c|c|c|c|}
\hline
  & & $F_+(0)$   & $V(0)$ & $A_1(0)$ & $A_2(0)$ & $V(0)/ A_1(0)$ &
$A_2(0)/A_1(0)$ \\
 \hline\hline
Lat      & \cite{lat3}a & 0.29$\pm$ 0.06  & 0.45$\pm$0.22 & 0.29$\pm$0.16 &
0.24$\pm$ 0.56 & 2.0$\pm$0.9 & 0.8$\pm$1.5 \\
         & \cite{lat3}b & 0.35$\pm$ 0.08  & 0.53$\pm$0.31 & 0.24$\pm$0.12 &
0.27$\pm$ 0.80 & 2.6$\pm$1.9 & 1.0$\pm$3.1 \\

         & \cite{lat4}a & 0.26$\pm$ 0.16  & 0.34$\pm$0.10 & 0.25$\pm$0.06 &
0.38$\pm$ 0.22 & 1.4$\pm$0.2 & 1.5$\pm$0.7 \\
         & \cite{lat4}b & 0.30$\pm$ 0.19  & 0.37$\pm$0.11 & 0.22$\pm$0.05 &
0.49$\pm$ 0.26 & 1.6$\pm$0.3 & 2.3$\pm$0.9 \\
\hline
SR       &\cite{sr1}    & 0.24$\pm$ 0.025 &    --         &     --        &
--         &      --     &     --      \\
         &\cite{sr3}    & 0.40$\pm$ 0.20  &    --         &     --        &
--         &      --     &     --      \\
\hline
QM     &WSB\cite{wsb}   &  0.33           &     0.33      &      0.28     &
0.28       &      1.2    &    1.0      \\
       & GISW\cite{gisw}&  0.09           &     0.27      &      0.05     &
0.02       &      5.4    &    0.4      \\
\hline\hline
\end{tabular}
\end{table}

\begin{table}[2]
\caption{\label{table:d2k}
The form factors of the decays $D\to K,K^*$ at $q^2=0$. }
\centering
\begin{tabular}{|c|c|c|c|c|c|c|c|}
\hline
         &              &   $F_+(0)$      &     $V(0)$    &     $A_1(0)$  &
$A_2(0)$    &$V(0)/ A_1(0)$&$A_2(0)/A_1(0)$\\
\hline\hline
Exp      & \cite{exp6}  & 0.77$\pm$ 0.04  & 1.16$\pm$0.16 & 0.61$\pm$0.05 &
0.45$\pm$ 0.09 & 1.90$\pm$0.25& 0.74$\pm$0.15 \\
\hline
Lat      & \cite{lat3}  & 0.78$\pm$ 0.08  & 1.08$\pm$0.22 & 0.67$\pm$0.11 &
0.49$\pm$ 0.34 & 1.6$\pm$0.3  & 0.7$\pm$0.4   \\

         & \cite{lat4}  & 0.60$\pm$ 0.22  & 0.86$\pm$0.24 & 0.64$\pm$0.16 &
0.40$\pm$ 0.32 & 1.3$\pm$0.2  & 0.6$\pm$0.3   \\
\hline
SR       &\cite{sr2}    & 0.6$\pm$ 0.15   & 1.1$\pm$ 0.25 & 0.5$\pm$ 0.15 &
0.6$\pm$ 0.1   & 2.2$\pm$0.2  &  1.2$\pm$ 0.2 \\
\hline
QM       & WSB\cite{wsb}&  0.76           &     1.23      &      0.88     &
1.15       &      1.4     &    1.3        \\
        &GISW\cite{gisw}&  0.8            &     1.10      &      0.80     &
0.80       &      1.4     &    1.0        \\
\hline
LCQM     & \cite{jaus}  &  0.73           &     0.92      &      0.63     &
0.42       &      1.46    &   0.67        \\
\hline\hline
\end{tabular}
\end{table}

\begin{table}[3]
\caption{\label{table:decconst}
The decay constants $f_P$ of pseudoscalar mesons $MeV$. }
\centering
\begin{tabular}{|c|c|c|c|c|}
\hline
                     &        $\pi$    &     $K$       &     $D$       &
$B$      \\
\hline\hline
Exp \cite{exp7}      & 130.7$\pm$ 0.46 & 159.8$\pm$1.9 &  $<$ 310      &
--       \\
\hline
Lattice \cite{lat6}  &                 &               &  200$\pm$ 30    &
180$\pm$ 40  \\
\hline
Sum Rules            &       --        &      --       &160 \cite{sr2} &    --
         \\
                     &       --        &      --       &165$\div$195\cite{sr5}&
130$\div$ 200\cite{sr5} \\
\hline
LCQM     \cite{card} &  130.7          &     162       &     220       &
188        \\
LCQM\cite{schlumpf}  &  130.7          &     162       &     206       &
186        \\
This work            &  130            &     160       &     234       &
202        \\
\hline\hline
\end{tabular}
\end{table}

\begin{table}[4]
\caption{\label{table:heavymeson}
The decay constants $f_P$ of pseudoscalar mesons built up of quarks
with the masses $m_Q$ and $m_q$ and the slope of $h_{el}$ at $\omega=1$
calculated from $<M_Q|\bar Q\gamma_\mu Q|M_Q>$ as functions of $m_Q$ at
$m_q=0.25\;GeV$.}
\centering
\begin{tabular}{|c||c|c||c|c||c|c||c|c|}
\hline
& \multicolumn{2}{c||} {$\alpha=0.01$} & \multicolumn{2}{c||} {$\alpha=0.02$} &
\multicolumn{2}{c||} {$\alpha=0.04$} & \multicolumn{2}{c|} {$\alpha=0.08$} \\
\hline
$m_Q, GeV$ & $f_P,MeV$ & $\rho^2_{el}$ &  $f_P,MeV$ & $\rho^2_{el}$ & $f_P,MeV$
& $\rho^2_{el}$ & $f_P,MeV$ &$\rho^2_{el}$ \\
\hline\hline
0.25 &   151 & 0.04 & 130  & 0.06 &  104 & 0.08 &  80 & 0.1  \\
\hline
0.4  &   190 & 0.25 & 160  & 0.35 &  128 & 0.5  &  97 & 0.65 \\
\hline
1.8  &   324 & 0.6  & 234  & 0.65 &  163 & 0.82  & 110 & 1.0  \\
\hline
5.2  &   308 & 0.75 & 202  & 1.0  &  132 & 1.05 &  85 & 1.1  \\
\hline
10   &   254 & 1.0  & 162  & 1.05 &  102 & 1.1  &  64 & 1.25 \\
\hline
20   &   195 & 1.0  & 122  & 1.1  &   76 & 1.23 &  48 & 1.45 \\
\hline
40   &   143 & 1.0  &  89  & 1.11 &   55 & 1.25 &  34 & 1.66 \\
\hline
80   &   103 & 1.0  &  63  & 1.11 &   39 & 1.25 &  24 & 1.66 \\
\hline
\end{tabular}
\end{table}

\begin{table}[5]
\caption{\label{table:parameters}
The constituent quark masses and the calculated $f_P$ for $\alpha=0.02$.}
\centering
\begin{tabular}{|c|c||c|c|c|}
\hline
quark &quark mass,$GeV$& meson &meson mass,$GeV$ &$f_P,MeV$  \\
\hline\hline
u,d   & 0.25   &$\pi^+(u\bar d)  $& 0.14       &    130      \\
\hline
s     & 0.40   &$K^+(u\bar s)    $& 0.49       &    160      \\
\hline
c     & 1.80   &$D^+(c\bar d)    $& 1.87       &    234      \\
\hline
b     & 5.20   &$B^+(u\bar b)    $& 5.27       &    202      \\
\hline
\end{tabular}
\end{table}

\begin{table}[6]
\caption{\label{table:fits}
The parameters of the monopole and dipole fits to the $F_+$ form factor.
The masses of the lowest vector mesons which are expected to dominate the
form factors are given in brackets. }
\centering
\begin{tabular}{|c||c|c|c||c|c|}
\hline
  & \multicolumn{3}{c||} {$\alpha_B=\alpha_D=0.02$} & \multicolumn{2}{c|}
{$\alpha_B=\alpha_D=0.04$}  \\
\hline
Decay & $F_+(0)$ & $M_{mon},\;GeV$  & $M_{dip},\;GeV$ & $F_+(0)$ &
$M_{dip},\;GeV$  \\
\hline\hline
$B\to D$   & 0.73 & 5.7           & 7.7  & 0.68 & 7.20  \\
\hline
$B\to \pi$ & 0.23 & 5.2 [5.324]   & 6.2  & 0.22 & 6.08  \\
\hline
$D\to K$   & 0.70 & 2.22 [2.11]   & 3.0  & 0.70 & 2.95  \\
\hline
$D\to\pi$  & 0.55 & 2.1 [2.01]    & 2.8  & 0.59 & 2.68  \\
\hline
\end{tabular}
\end{table}

\begin{table}[7]
\caption{\label{table:rho2}
The slope of the IW function $\rho^2$}
\centering
\begin{tabular}{|c|c|c|c|c|c|c|}
\hline
 & ARGUS \cite{iwexp1} & CLEO \cite{iwexp2}&Lat
\cite{lat5}&SR\cite{sr6}&SR\cite{sr7}&This work \\
\hline
$\rho^2$  &  $1.07\pm 0.17$ &  $0.87\pm 0.12$      &1.2            &$>1.04$
 & $0.7\pm0.2$  &
$0.8\pm0.1$   \\
\hline
\end{tabular}
\end{table}

\begin{table}[8]
\caption{\label{table:iw}
The parameters of the IW function.}
\centering
\begin{tabular}{|c|c||c|c|}
\hline
$\alpha_D$  & $\alpha_B $ & $\xi(1)$   & $\rho^2$  \\
\hline
0.02  &  0.02  & 0.98   &  0.78   \\
\hline
0.02  &  0.04  & 0.87   &  0.75   \\
\hline
0.04  &  0.02  & 0.93   &  0.7   \\
\hline
0.04  &  0.04  & 0.98   &  0.88   \\
\hline\hline
\end{tabular}
\end{table}

\section{Appendix A: Bound state description within dispersion relations}
\begin{figure}
\begin{center}  \mbox{   \epsfig{file=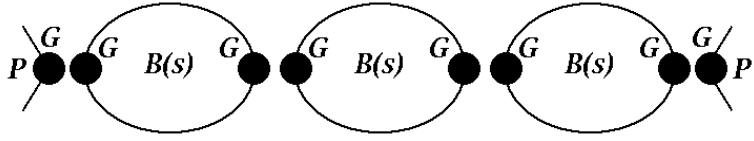,height=2cm}    }
\end{center}
\caption{ One of the terms in the expansion of $A_0(s)$
\label{fig:a1}}
\end{figure}

To illustrate main points of the dispersion approach we
consider the case of two spinless constituents with the masses $m_1$ and $m_2$
interacting via
exchanges of a meson with the mass $\mu$. We start with the scattering
amplitude
\begin{equation}
A(s,t)=<k'_1,k'_2|S|k_1,k_2>, \quad s=(k_1+k_2)^2,\; t=(k_1-k'_1)^2
\end{equation}
The amplitude as a function of $s$ has the
threshold singularities in the complex $s$-plane connected with
elastic rescatterings of the constituents and production of new mesons at
\begin{equation}
s= (m_1+m_2)^2,\;(m_1+m_2+\mu)^2,\;(m_1+m_2+2\mu)^2\ldots
\end{equation}
We assume that an $S$-wave bound state with the mass $M<m_1+m_2$ exists, then
the partial amplitude $A_0(s)$ has a pole at $s=M^2$.
The amplitude $A(s,t)$ has also $t$-channel singularities at
$t=(n\mu)^2;\;\; n=1,2,3\ldots$ connected with meson exchanges.
If one needs to construct the amplitude in the low-energy region $s\geq
(m_1+m_2)^2$
the dispersion $N/D$ representation turns out to be convenient.
Consider the $S$-wave partial amplitude
\begin{equation}
A_0(s)=\int\limits^1_{-1} dz\, A(s,t(s,z)),
\end{equation}
where
$t(z)=-(1-z)\lambda(s,m_1^2,m_2^2)/2s$, $z=\cos\theta$ in the c.m.s.
The $A_0(s)$ as a function of complex $s$ has the right-hand
singularities related to $s$-channel singularities of $A(s,t)$.
In addition, it has left-hand singularities located at
$s=(m_1+m_2)^2-(n\mu)^2;\;\; n=1,2,3\ldots$. They come from $t$-channel
singularities
of $A(s,t)$.
The unitarity condition in the region $s\approx (m_1+m_2)^2$ reads
\begin{equation}
{\rm Im}\, A_0(s)= \rho(s)\;|A_0(s)|^2,\qquad
\rho(s) =\frac{\lambda(s,m_1^2,m_2^2)}{16\pi s}
\end{equation}
with
$\rho(s)$ the two-particle phase space.
The $N/D$ method represents the partial amplitude as $A_0(s)=N(s)/D(s)$, where
the function $N$ has only left-hand singularities and $D$ has only right-hand
ones. The unitarity condition yields
\begin{equation}
D(s) = 1 - \int\limits^\infty_{(m_1+m_2)^2}\frac{d\tilde{s}}{\pi}\,
\frac{\rho(\tilde{s})N
(\tilde{s})}{\tilde{s}- s}\; \equiv \; 1-B(s).
\end{equation}
Assuming the function $N$ to be positive we introduce
$G(s)=\sqrt{N(s)}$.
Then the partial amplitude takes the form
\begin{equation}
A_0(s)=G(s)\left[1+B(s)+B^2(s)+B^3(s)+\ldots\right] G(s)
=\frac{G(s)G(s)}{1-B(s)}.
\end{equation}
This expression can be interpreted as a series of loop diagrams of
Fig.\ref{fig:a1}
with the basic loop diagram
\begin{equation}
B(s)=\int\limits^\infty_{(m_1+m_2)^2}\frac{d\tilde{s}}{\pi}\,
\frac{\rho(\tilde{s})\;
G^2(\tilde{s})}{\tilde{s}- s}.
\end{equation}
The bound state with the mass $M$ relates to a pole both in the total
and partial amplitudes at $s=M^2$ so $B(M^2)=1$.
Near the pole one has for the total amplitude
\begin{eqnarray}
A&=&<k'_1,k'_2|P>\frac1{M^2-P^2}<P|k_1,k_2>+{\rm regular\; terms} \nonumber \\
&\equiv&\chi^*_P(k'_1,k'_2)\frac1{M^2-P^2}\chi_P(k_1,k_2)+\ldots
\end{eqnarray}
where $\chi_P(k_1,k_2)$ is the amputated Bethe-Salpeter amplitude
of the bound state.
The dispersion amplitude near the pole reads
$$
A=N/D+{\rm regular\; terms\; related\; to\; other\; partial\; waves}
$$
\begin{equation}
=\frac{G^2(M^2)}{(M^2-s)B'(M^2)}+\ldots
\equiv \frac{G_v^2(M^2)}{M^2-s}+\ldots
\end{equation}
where $G_v$ is a vertex of the bound state transition to the constituents.
The singular terms correspond to each other and hence
\begin{equation}
\chi_P(k_1,k_2)\to G_v(P^2)\equiv \frac{G(P^2)}{\sqrt{B'(M^2)}} \label{bsa}
\end{equation}
Underline that among right-hand singularities the constructed dispersion
amplitude takes into account only the two-particle cut.

Let us turn to the interaction of the two-constituent system with an external
electromagnetic field. The amplitude of this process
$T_\mu=<k'_1,k'_2|J_\mu(q)|k_1,k_2>$ in the case of a bound state takes the
form
\begin{eqnarray}
T_\mu&=& <k'_1,k'_2|P'>
\frac1{P'^2-M^2}
<P'|J_\mu(q)|P>
\frac1{P^2-M^2}
<P|k_1,k_2>+\ldots  \label{tmu} \nonumber \\
&=&\chi^*_P(k'_1,k'_2)
\frac1{P'^2-M^2}
(P'+P)_\mu F(q^2)
\frac1{P^2-M^2}
\chi_P(k_1,k_2)+\ldots
\end{eqnarray}
where the bound state form factor  is defined as
\begin{equation}
<P'|J_\mu(q)|P>=(P'+P)_\mu F(q^2)
\end{equation}

\begin{figure}
\begin{center}\mbox{\epsfig{file=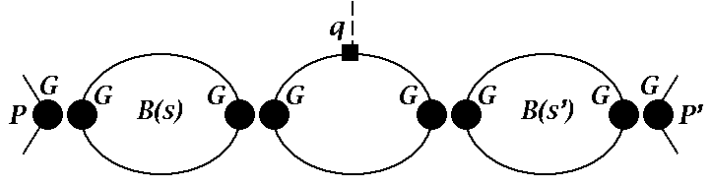,height=2.5cm}}
\end{center}
\caption{ One of the terms in the series for $T_\mu$.
\label{fig:a2}}
\end{figure}

The dispersion amplitude
$T_\mu$ with only two-particle singularities in the $P^2$- and
$P'^2$-channels taken into account is given \cite{akms} by the series of graphs
in Fig.\ref{fig:a2}.

These graphs are obtained from the dispersion scattering
amplitude series by inserting a photon line into constituent lines.
The amplitude reads
\begin{equation}
T_\mu(P',P,q)=2P_\mu(q)T(s',s,q^2)+\frac{q_\mu}{q^2}C,
\end{equation}
$$
P^2=s,\;P'^2=s',\;q=P'-P,\;P_\mu(q)=(P-\frac{qP}{q^2}\,q)_\mu
$$
The dispersion method allows one to determine
$T(s,s',q^2)$, which is the part of the amplitude transverse with respect
to $q_\mu$.
Summing up the series of dispersion graphs in Fig.\ref{fig:2} gives
\begin{equation}
T(s',s,q^2)=\frac{G(s)}{1-B(s)}\Gamma(s',s,q^2)\frac{G(s')}{1-B(s')}.
\end{equation}
Here
$$
\Gamma(s',s,q^2)=\int\frac{d\tilde{s}G(\tilde{s})}{\pi(\tilde{s}-s)}
\frac{d\tilde{s}'
G(\tilde{s}')}{\pi(\tilde{s}'-s)}\Delta(\tilde{s}',\tilde{s},q^2),
$$
and $\Delta(\tilde{s}',\tilde{s},q^2)$
is the double spectral density of the three-point Feynman graph with a
pointlike vertex of the constituent interaction.

The longitudinal part $C$ is given by the Ward identity
\begin{equation}
C=\frac{G(s)}{1-B(s)}\left( {B(s')-B(s)} \right) \frac{G(s')}{1-B(s')}
\end{equation}

At $s=s'=M^2$, the quantity $T_\mu$ develops both $s$ and $s'$ poles, so
\begin{equation}
\label{ff}
T_\mu(P',P,q)=\frac{G_v(M^2)}{M^2-s}(P'+P)_\mu F(q^2)
\frac{G_v(M^2)}{M^2-s'}+{\rm less\;singular\;terms}
\end{equation}
where
\begin{equation}
\label{ffv}
F(q^2)=\int\limits_{(m_1+m_2)^2}^\infty\frac{ds G_v (s)}{\pi(s -M^2)}
\frac{ds' G_v (s')}{\pi(s'-M^2)}\Delta(s',s,q^2).
\end{equation}
is the bound--state form factor  (see (\ref{bsa}) and (\ref{tmu})).
So, the quantity $<P'|J_\mu(q)|P>$ corresponds to the three--point dispersion
graph
with the vertices $G_v$.
The following relation is valid
$\Delta(s',s,0)=\pi\delta(s'-s)\rho(s)$.
This is a consequence of the Ward identity which relates the
three-point graph at zero
momentum transfer to the loop graph. This relation yields the charge
normalization $F(0)=1$. The expression (\ref{ffv}) gives the form factor  in
terms of
the $N$-function of the constituent scattering amplitude and double
spectral density of the Feynman graph.
In general, the following prescription works: to obtain the dispersion
expression
spectral density in channels corresponding to a bound state, one should
calculate
the related Feynman graph spectral density and multiply it by $G_v$.

If the constituent is a nonpoint particle,
the expression (\ref{ffv}) should be multiplied by form factor  of an on-shell
constituent.

\end{document}